\documentclass[conference]{IEEEtran}
\usepackage{graphicx}
\usepackage[table]{xcolor}
\usepackage{makecell}
\usepackage[export]{adjustbox}
\usepackage{tabularx}
\usepackage{booktabs}

\ifCLASSINFOpdf
\else
\fi
\begin{document}
%
\title{Building Persuasive Robots with Social Power Strategies}

\author{
\IEEEauthorblockN{Mojgan Hashemian\IEEEauthorrefmark{1}\IEEEauthorrefmark{2}, Marta Couto\IEEEauthorrefmark{1}, Samuel Mascarenhas\IEEEauthorrefmark{1}\IEEEauthorrefmark{2}, Ana Paiva\IEEEauthorrefmark{1}\IEEEauthorrefmark{2}, Pedro A. Santos\IEEEauthorrefmark{1}\IEEEauthorrefmark{2},  Rui Prada\IEEEauthorrefmark{1}\IEEEauthorrefmark{2}
}
\IEEEauthorblockA{\IEEEauthorrefmark{1}INESC-ID}
\IEEEauthorblockA{\IEEEauthorrefmark{2}
Instituto Superior Técnico, Universisade de Lisboa}
}
\maketitle

\begin{abstract}
Can social power endow social robots with the capacity to persuade? This paper represents our recent endeavor to design persuasive social robots. We have designed and run three different user studies to investigate the effectiveness of different bases of social power (inspired by French and Raven's theory) on peoples' compliance to the requests of social robots. 
The results show that robotic persuaders that exert social power (specifically from expert, reward, and coercion bases) demonstrate increased ability to influence humans.
The first study provides a positive answer and shows that under the same circumstances, people with different personalities prefer robots using a specific social power base. In addition, social rewards can be useful in persuading individuals. The second study suggests that by employing social power, social robots are capable of persuading people objectively to select a less desirable choice among others. Finally, the third study shows that the effect of power on persuasion does not decay over time and might strengthen under specific circumstances. Moreover, exerting stronger social power does not necessarily lead to higher persuasion. 
Overall, we argue that the results of these studies are relevant for designing human--robot-interaction scenarios especially the ones aiming at behavioral change. 

\end{abstract}


%
\IEEEpeerreviewmaketitle


\section{Introduction}
Social power, i.e., the potential for social influence, is a pervasive social process in human--human interactions. However, despite its acknowledged role in social interaction, little attention has been paid to this phenomenon in human--robot interaction (HRI). 
One prominent example of social agents that have been of interest these days is robotic agents' evolution. However, concerning robotic agents, few studies have addressed social power in HRI.

Recent advances in the field of social robotics raise the question that whether a social robot can be used as a persuasive agent. To date, few attempts have been performed using different approaches to tackle this research question (reviewed in Section~\ref{rw}). 
In a nutshell, the present work's objective is to empower intelligent agents with social power dynamics that are aiming to develop more persuasive agents. In this paper, we report our recent advancements and draw suggestions for future directions.

Recent evidence suggests that social power (or, in short, power) is central to a multitude of social processes. Power acts as a \textit{heuristic} solution to potential conflicts among group members and guides social perception and behavior~\cite{keltner2008reciprocal}. Extensive research in the field of social psychology has shown that social power affects a wide variety of social and cognitive processes, such as stereotyping~\cite{fiske1993controlling}, moral judgment~\cite{fiske1992four} as well as  nonverbal behavior, like emotional displays~\cite{clark1990emotions}, and its inferences~\cite{hall1994subordination}.

Additionally, recent evidence suggests that humans perceive computers as social agents, and people respond socially to computer actors (computers are Social actors [CASA] paradigm). In other words, humans treat computers similarly to how they treat other humans~\cite{nass1993anthropomorphism}. In this sense, people apply similar social rules to their relationship with computers~\cite{reeves1996media}. The same might apply to social power theories for computers.  

We argue that more research on social power dynamics is needed to create socially competent robots. The studies presented here contribute to this. We approach this goal from two perspectives, since any power-related relationship deals with two sides: the agent who exerts power (the actor) and the target.
%
Specifically, we aim to design social robots capable of 1) processing social power dynamics, 2) representing power in their behavior, and 3) investigating how they are perceived when using different power sources. 
To operationalize the expression of power, we propose to utilize persuasion as an application of social power. 

In sum, we aim to investigate the link between social power and persuasion in social robotics. Specifically, we would like to understand how to design more persuasive robots using social power. 
Furthermore, by operationalizing social power in the context of persuasion, we develop different persuasion studies based on three different power bases. We investigate the effectiveness of these persuasive strategies by designing and implementing three different user studies.

This paper is organized as follows. Section~\ref{bg} introduces the terminology and definition of social power and persuasion. A brief overview of the recent advancements in the field of persuasive social robots is presented in Section~\ref{rw}.
The paper's remaining sections describe our recent endeavors in designing persuasive social robots using social power. 
Section~\ref{method} details our studies' general methodology, followed by Section \ref{study1}, which details the results obtained in the first user study. Then, Section~\ref{study2} presents the findings of the second study,
and Section~\ref{study3} begins by laying out the details of the third study, followed by the findings.
Finally, the last Section displays our conclusions, focusing on the two key themes that highlight the studies' limitations and providing recommendations for further research work.

\section{Background}~\label{bg}
Social power is defined as the ability to influence others to do something they would not do in the absence of such power~\cite{pierro2008motivated}. 
This study uses a well-known theory of social power introduced by French and Raven~\cite{french1959bases}, which presents one of the most popular models of social power; many researchers have tested this model within the context of many applications. Further, its widely accepted conceptualizations have been examined in terms of applicability to various settings~\cite{elias2008fifty}.  Moreover, the model has been verified after 30 years, which offers a good reliability base~\cite{raven1988french}.   

The authors' model considers five principal bases for social power: reward, coercion, legitimacy, expertise, and reference. The model is based on a typological analysis of the bases of power in interpersonal influence, making it immensely interesting for our second goal of designing persuasive agents.

French and Raven \cite{french1959bases}  identified different bases of power, i.e., resources, that can change another person's beliefs, behaviors, and attitude. Although there have been other identified bases, in \cite{french1959bases}, the authors argue that these five bases are the most common ones among the others. The definitions of the five bases are as follows:
\begin{itemize}
    \item \textbf{Reward} social power is realized when the target is willing to do what the actor requests in response to another action that brings value to the target in their perspective. For example, a factory manager (the actor) promises an employee (the target) to double their salary if they increase production. 
    \item \textbf{Coercive} power stems from the ability of one individual to mediate punishments for another. For example, a factory manager (the actor) warns a worker (the target) that if they do not increase production, they will be fired.
    \item \textbf{Legitimate} power stems from internalized values that give one individual the authority to influence another. For example, in a family, a parent (the actor) instructs their teenage child (the target) to be home before midnight. In case of a successful interaction, the teenager gets home earlier, since, generally, children recognize their parent's authority to enforce a curfew.
    
    \item \textbf{Expert} power stems from one individual's perception of other's higher knowledge. For example, a physician (the actor) instructs a patient (the target) to follow a given medical prescription. 
    \item \textbf{Referent} power stems from liking, respect, and \textit{identification} of one individual with another. For example, a person (the actor) asks a friend (he target) for help in studying for an upcoming exam. Another example of referent power could be the influence of celebrities or social media influencers on people's decisions.
\end{itemize}

One way of exerting social power is to persuade people. \textit{Persuasion} is defined as an attempt to change/shape a target's belief or behavior about a subject, an issue, or an object~\cite{siegel2009persuasive, fogg2002persuasive}. Hence, persuasion involves the study of attitudes and how to change them~\cite{perloff1993dynamics}. In other words, persuasion may be defined as attitude change formation through information processing in response to a message about an intended object ~\cite{bohner2008information}.

In this process, by transmitting a message, the communicator tries to convince others to change their attitude or behavior regarding an issue, in an atmosphere of \textit{free choice}. In other words, the persuader conveys a message, which aims at convincing other people to change their attitude; this is usually not forced upon people. 

Here, it is noteworthy that the communicator is not changing people's minds. Conversely, in case of a successful persuasion, the target changes their attitudes. However, of course, persuasion is not always successful. Sometimes, the target can react negatively to persuasion attempts and actively avoid being persuaded. This phenomenon is called reactance~\cite{perloff1993dynamics}. Given this context, \textit{persuadability} is not an individual characteristic but rather a ``complex communication phenomenon." 


Persuasion is a key process in shaping and maintaining cooperation, social influence, and behavioral change~\cite{reardon1991persuasion}. It plays a critical role in human interaction and exchanges~\cite{oreg2014source}, and several factors contribute to its effectiveness, such as the personality of the actor (the source or the one who is performing the influence) and the target (the one who is affected)~\cite{oreg2014source, anagnostopoulou2017exploring}. 
It should be noted that there is contradicting evidence regarding personality. For instance, some psychologists believe that no personality trait is associated with persuasion due to the complexity of human behavior~\cite{perloff1993dynamics}.

To understand the process of \textit{being persuaded}, it is essential to comprehend the target's perception of the persuader's characteristics (e.g., the target's internal cognitive process). On the contrary, in understanding the process of \textit{persuading}, the actor's characteristics play a vital role (e.g., actions of the actor).

Social power and persuasion have a special bond that motivated us to investigate them jointly in the context of social robotics. If one considers the definition of social power as the ability to influence others, this means that the relationship between social power and persuasion is already established. In fact, they are almost inseparable. 
Additionally, persuasion is ``an important medium of social power"~\cite{dowding2011encyclopedia}, which motivated us to investigate the effect of social power dynamics on social agents' persuasiveness and its potential effect on influencing others. Thus, persuasion attempts to change/shape a target's belief or behavior about a subject, an issue, or an object~\cite{fogg2002persuasive}.

In the field of social psychology, the link between power and persuasion has been the subject of investigation for a long time~\cite{cutlip1960power} (for a recent review, see \cite{brinol2017power}). 
Early results show that a powerful individual is more influential in persuading others~\cite{hovland1949experiments}. However, it should be noted that the extent to which the power is effective depends on the circumstances by which it can cause short-/long-term influence and increase or decrease persuasion~\cite{brinol2017power}.

%
Specifically, some theories indicate a linear correlation between power and persuasion. In other words, when more power is exerted, higher persuasion is achieved. However, recent evidence argues that this is generally not true. Under specific circumstances, when higher power is exerted, reactance comes into play and decreases the chance of persuasion. This happens because the persuasiveness of messages depends on the psychological sense of power. Hence, a high-power communicator may lead to high or low persuasion depending on the power level of the audience (the persuasion target).

For instance, evidence suggests that during mock interviews, when the power levels of the interviewer and the interviewee match, higher persuasion is achieved~\cite{dubois2016dynamics}. In other words, high-power communicators are more effective in persuading high-power audiences; similarly, low-power communicators are more effective in persuading low-power audiences. As an example, when both interviewer and interviewee are in low-power state, the interviewer finds the target more persuasive. This contradicts earlier studies that stated interviewees with high power are more persuasive~\cite{lammers2013power}. Recent finding suggests that this inconsistency in the results is due to the mismatch between the powers. In other words, this inconsistency is observed when low-power people are interviewed by low-power interviewers. Hence, the persuasiveness of messages depends on the psychological sense of power of the two sides.

Additionally, high-power people generate and pay greater emphasis on information, which conveys competence (e.g., by stressing on skillfulness and intelligence). 
On the contrary, a low-power state leads to more warmth, i.e., low-power communicators generate messages with more warmth, for instance, by stressing on friendliness and trustworthiness~\cite{dubois2016dynamics}. This motivated us to investigate different persuasive strategies in terms of information, competence, and warmth (discussed in the following sections). 

Having discussed the terminologies and the link between social power and persuasion, in the next section, we discuss the recent advancements in the field of persuasive social robots in HRI.


\section{Related Work}~\label{rw}


To date, a considerable amount of literature has been published on persuasive robots, but few studies have considered social power in persuasion. Several lines of evidence suggest that robots can be used as persuasive interlocutors. Currently, much of the existing literature pays particular attention to behavioral strategies and nonverbal cues, either social (such as mimicry~\cite{ghazali2019assessing}) or physical (such as gender~\cite{siegel2009persuasive}, embodiment~\cite{li2015benefit, herse2018bon}, gaze~\cite{rossi2017gaze}). 
Additionally, previous studies reveal several factors associated with the ability of an individual to persuade others. These factors include verbal and nonverbal behaviors of the individual, the dynamics of social interaction, and psychological and societal factors, such as social roles~\cite{chidambaram2012designing}.

We would like to highlight that so far, little attention has been paid to the importance of \textit{message strategy}, or the way that a robot phrases a request appeal to gain higher compliance. We briefly review these studies in this section.

In a recent research~\cite{saunderson2019would}, ten multi-modal persuasive strategies (direct request, cooperation, critique, threat, deceit, liking, logic, affect, exclusivity, and authority appeal) were selected and coded verbally; these were combined with specific gestures validated in a pilot study. A study was conducted with 200 people, who played the jelly bean game (involves visually estimating the number of specific jelly beans in a bottle). Prior to their guessing, the robots gave their suggestions and attempted to influence the users' decision. The task was performed with two NAO robots, and unique strategies were randomly assigned to each robot. The results show that affect and logic strategy gained the most compliance. 

A followup study~\cite{saunderson2020investigating} using the same game investigates further the effect of affect and logic strategy. This study includes a control condition. Specifically, one robot with no strategy was added as a control condition and compared with another robot that was equipped with a persuasive strategy.
The control robot stated neutral messages (e.g., ``There are $x$ number of jelly beans in the jar") gestures (standing).
The persuasive strategy-equipped robot used verbal cues to persuade the user (e.g, in emotional condition, the robot stated, ``It would make me happy if you use my guess of $x$ beans in the jar," and in the logic condition, the robot stated, ``My computer vision system can detect $x$ number of beans in the jar.").
The results indicate that the emotional strategy was more persuasive than the logic and control condition. No statistically significant difference was found between the logic and the control condition.

Another interesting work investigates the effect of foot-in-the-door (FITD) technique, which starts by a small and moderate requests that a person accepts and then continues to get a person to agree with a larger request~\cite{lee2019robotic} . To be more specific, the robot attempts to persuade the user using the sequential-request strategy, starting from an easy one. 
The authors argue that persuasiveness might depend on the performance and credibility of robots.
Bearing this in mind, the authors ran a user study with 44 people in four conditions: 2 (robot performance: helpful vs. unhelpful) $\times$ 2 (message strategy: direct request vs. foot-in-the-door).
The results indicate that this technique can be used by robots to persuade human users. However, the persuasion effect was independent of the robot's expertise and credibility.

Using a similar approach, in \cite{kobberholm2020influence}, the authors attempt to investigate the effect of incremental representation of information on persuasiveness of social robots. In a between-subject study with two conditions of incremental and non-incremental information presentation, the NAO robot tried to persuade the users to do a higher number of  tasks (ten simple tasks in total). The tasks used in the two conditions were the same. In the non-incremental condition, the information about all the tasks was given at once, while in the incremental condition, the participants received the information when the next task was about to start. The result did not yield any significant differences regarding the number of the task and the likeability of the robot. However, the participants were persuaded to stay longer to do the tasks after they had intended to leave.

In addition, Andrist et al.~\cite{andrist2013rhetorical} studied the effects of rhetorical ability in expertise communication of informational robots using psychological and linguistic theories. They ran a study with 44 participants, using two Lego Mindstorm robots in four conditions by expressing four types of expertise by means of low/high practical knowledge and low/high rhetorical ability. The robots employed linguistic cues in their speech to provide expertise effectively in order to raise trust and gain compliance. Each robot expressed different levels of expertise and rhetorical ability depending on the condition. To express linguistic ability, the robot used any one of the five following linguistic cues: goodwill, prior expertise, organization, metaphor, and fluency. 
The results indicate that the speech using linguistic cues was more effective than the one with practical knowledge and simple facts. 
Thus, the increase in linguistic cues leads to higher persuasion.

In sum, although some research studies have already explored how robots can be more persuasive, we aim to study how social power can also be used by social robots as a persuasion mechanism. Our approach to this research question is discussed in the following sections.

\section{Methodology}~\label{method}
We designed and performed three user studies to investigate how different bases of social power contribute to the persuasiveness of social robots. 

In Study 1, we programmed two robots using either expert or reward social power in an adversarial setting. In other words, we programmed one robot to express expertise by giving information to the users. Additionally, as a specific instance of reward, we used social rewards (``telling a joke" by the other robot). This study is further discussed in Section~\ref{study1}.

In Study 2, a single robot used two different strategies to persuade the user (reward and coercion), and the results were compared to that of the control condition. Within the three conditions (reward, coercion, and control), the robot attempted to persuade the users to select a less desirable choice among others. Section~\ref{study2} presents this second study.

Last, in Study 3, a robot used one power strategy (reward) but with different strengths, and the obtained results were compared with those of two control conditions: with and without the presence of any robot. In this study, the persuasion attempt was repeated over a series of interactions. Details of this study are presented in Section~\ref{study3}.

Overall, the results of the three user studies endorse that social power (in particular reward, coercion, and expert bases) endows persuasiveness to social robots and that different persuasive strategies could be perceived and preferred differently depending on users’ profiles and the persuasion context.

\section{Study 1}\label{study1}

\begin{figure}[!t]
\centering
\includegraphics[width=0.9\linewidth]{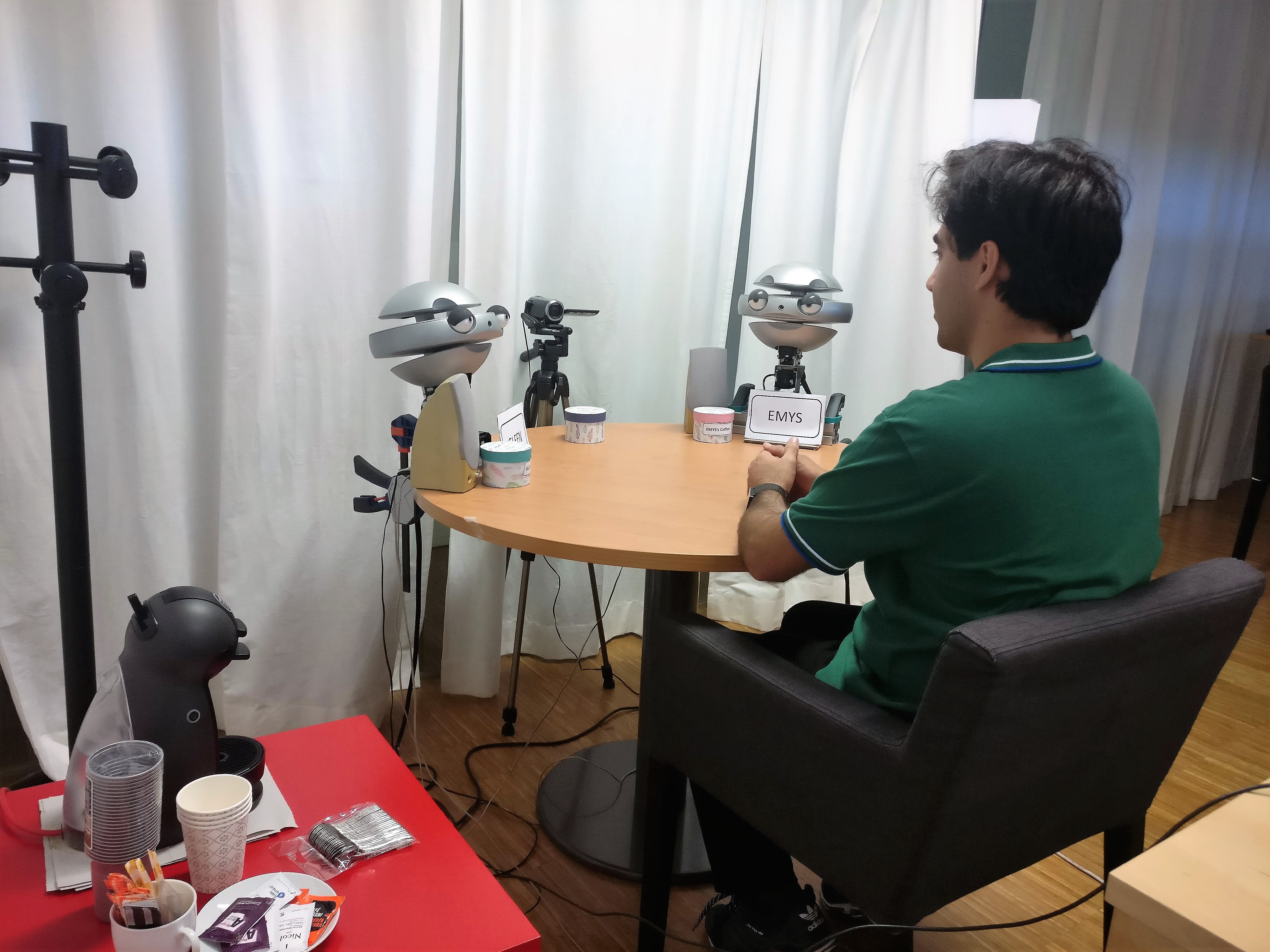}
\caption{Study 1: Setup.}
\label{fig:st1_setup}
\end{figure}

The first study is designed to achieve two main goals: understanding how social power makes the robots more persuasive and learning how different sources of social power lead to different user perception. (Fig.~\ref{fig:st1_setup} represents the setup of the study.)

Particularly, we aim to investigate the effect of reward and expert power strategies on the persuasibility of social robots. 
With this aim, to operationalize persuasive attitudes of robots, we employ these two strategies that are inspired by two different social power bases, i.e., reward and expertise~\cite{french1959bases}. 
That is to say, we design persuasive strategies inspired by these two sources of power, which from now on we refer to as reward/expert persuasive strategies. 
In so doing, we assign the role of an actor to robots and investigate their persuasiveness based on the specific power strategy in use.

We built the reward strategy using \textit{social rewards}. As a matter of fact, social interaction is rewarding for social species and can drive an individual's behavior~\cite{freeman2018effect}. Assuming the CASA paradigm~\cite{nass1994computers}, robots are perceived as social beings; hence, social rewards from robots would positively affect users' mental systems in a similar way~\cite{okumura2017social}.
In this context, in~\cite{wang2017development}, the results show that among children and adolescents, tangible social reward has stronger incentive power than monetary reward.
In addition, we argue that social rewards, unlike material rewards, could be unlimited and are, therefore, always available.

The concept of using social reward is not new and has already been used in a number of recent studies. For instance, positive facial expressions, such as smile and admiration, have been used in prior studies targeting children or adolescents \cite{wang2017development, kohls2009differential,demurie2012effects}. Inspired by such investigations on human--human interaction, recent studies on HRI have investigated the role of social rewards. For instance, in~\cite{okumura2017social}, the authors investigate the relationship between the effects of social rewards and offline improvements on motor skills. The results show that people who received the social reward performed better in the sequential finger-tapping task and that higher degree of satisfaction toward the robot's speech is achieved when social rewards were applied. In~\cite{midden2009using}, the social feedback was observed to have a stronger effect than factual feedback in persuading human users.

In this study, we use ``telling a joke'' as a social reward. 
Recently, researchers have shown an increased interest in humor in human--computer interaction (HCI)/HRI. Previous studies have investigated the concept of humor and telling a joke using computers or robots~\cite{bechade2016empirical, nijholt2016smart, valitutti2016infusing, khooshabeh2011does, kang2017social, vilk2020comedy}. 
Overall, these studies indicate that humor and jokes can modify the relationship and positive affect. Hence, we argue that telling a joke would be rewarding in a similar manner as other social rewards. 

According to the elaboration likelihood model (ELM)~\cite{petty1986elaboration}, there are two major routes to persuasion: \textit{central route}, in which the persuasive message is relevant to the persuadee and the quality of the arguments has an influence on attitudes;
\textit{peripheral route}, in which the persuasive messages are less relevant to the persuadee and the expertise of the source influences the change in the attitudes.   
Considering the dual process of persuasion, humor is considered to persuade via the peripheral route~\cite{gass2015persuasion}.

The other power base that we employ in this experiment is expert social power. Although robots, in general, hold  a great  potential as  informational  assistants, they must use an expert language to shape how helpful they are perceived by human users~\cite{torrey2009robots}. This is because as informational assistants, people expect  them  to  be  experts  in  their  area  of  specialty~\cite{andrist2013rhetorical}. In this direction, a number of recent studies have investigated different factors that can effect representation of expertise by informational social robots. 

For instance, in~\cite{torrey2006effects}, the authors investigate the degree to which an expert robot needs to represent information depending on the expertise level of the user. Specifically, they state that presenting too much information by a robot to a person who is an expert in that field might be rude or presenting too little information to a person who has no clue about a subject might be confusing or misleading. 
In a different study~\cite{torrey2009robots}, the same authors insinuate that softening the conversation by using expressions such as ``I think,'' ``maybe,'' and ``so forth'' might lead to a more polite robot.
Further, Andrist et al.~\cite{andrist2013rhetorical} claim that by using simple facts and rhetorical cues, robots can be perceived as experts in the targeted field.
In this study, we use a number of discrete facts and goodwill rhetorical abilities to design an expert social robot. 


Since the detailed design and main results of this study are already reported in~\cite{hashemian2019power}, here we just present a summary of the finding. 
At the end of the experiment, 51 people (17 females, 34 males) participate in the experiment voluntarily. The participants' age ranges from 20 to 55 years, with a mean of 29.45 $\pm$ 6.4. 

The study follows a within-subject design. Two EMYS robots (one called Emys and the other Gleen) promote one coffee capsule each. We add a third coffee option to control for random choice. One robot attempts to persuade the users using expert power by presenting information about the coffee (we call it the Expert). The other robot tries to persuade the users by giving them social rewards, i.e., by telling a joke (we refer to this robot as the Joker). The third coffee is not promoted by any of the robots and represents the control condition.
In sum, in a competitive scenario, one robot plays the role of an expert and the other tries to influence the user by giving them a social reward. They both try to persuade the user to select their own coffee brand.

We measure participants' personality and coffee drinking habit (CDH), i.e., how much they like coffee/how much coffee they drink, before interaction. Then, we record their coffee selection (which coffee they select), robot preference (which robot they prefer to interact with in general), perceived persuasiveness of robots (how persuasive they find each robot with a specific power strategy), robot perception (how they perceive each robot in terms of warmth, competence, and discomfort---using the Robotic Social Attribute Scale [RoSAS] questionnaire~\cite{carpinella2017robotic}), and future compliance (FC) toward robot (the likelihood of following the robot's suggestions in the future). 

In this study, we construct the following hypotheses:

\begin{itemize}
    \item \textbf{H1}: 
    The expert persuasive strategy would be more \textit{effective} than reward.
    \item \textbf{H2}: 
    The robot using a reward power strategy would be \textit{preferred} more than the one using an expert strategy.
    \item \textbf{H3}: 
    Reward increases the \textit{warmth} score of the robot and expertise increases the \textit{competence} score.
    \item \textbf{H4}: 
    The robot uses an expert strategy to be perceived more \textit{persuasive}.
    \item \textbf{H5}: 
    People would be more \textit{compliant} with the expert robot in the near future.
    \item \textbf{H6}: 
    Perceived persuasiveness of expert or reward strategy is dependent on participants' \textit{personality} traits.
\end{itemize}

\subsection{Results}

Since the study was performed in English and with non-native English speakers, we ensured that the participants understood the robots' dialogues. We asked them to rate on a 5-point Likert scale the extent to which they understood each of the robots' speech (``Please specify how much you perceived EMYS/Gleen's speech: I understood Emys/Gleen … [1] never--[5] all the time'').
The results indicated that majority of the participants (31 out of 51) fully understood the robots' speech (Option 4/5), 12 people understood moderately (Option 3), minority of people (8) had basic understanding (Option 2), and no one reported never understanding the robots (Option 1). Overall, the robots were understood 4.16 and 4.25 out of five times on the average. The results of the t-test indicated that these two scores are significantly higher than 3 or the mid-score (Emys: M = 4.16, SD = .967, S.E. = .135, t[50] = 8.544, p = .000; Gleen: M = 4.25, SD = .913, S.E. = .128, t[50] = 9.815, p = .000). 

The scores of RoSAS questionnaire revealed that the Joker succeeded in presenting itself as more friendly, since it scored significantly higher on warmth (Z = -4.409, p = .000). Conversely, the Expert succeeded in proving itself as more knowledgeable, skilled, and informative, since it scored higher on competence (Z = -4.286, p = .000). 
Also, since none of the two robots performed any manipulations on the discomfort dimension, no differences were observed between them in this regard (Z = -.199, p = .842). 
In other words, in this design, none of the robots showed signals of aggressiveness, danger, etc., on the discomfort dimension but rather expressed either competence (knowledgeable, intelligence, etc.) or warmth (sociable, kind, etc.). The preceding statements acknowledge the effect of our manipulations on the participants. 

Considering the perceived persuasiveness of the two robots, on the one hand, no statistically significant differences were found between the Joker and the Expert (Z = -.944, p = .345); 
on the other hand, the third coffee (the control option) was selected much less frequently (by 8 out of 51 people). Hence, the robots were able to perform some persuasion.
Together, these two findings indicated that the two power strategies are effective and the two robots were able to persuade people, although they were perceived differently with regard to competence and warmth. It should be noted that the persuasiveness mean score corresponding to the two robots is higher than the medium score (3.4 for the Joker and 3.6 for the Expert), which endorses their ability to persuade and influence the participants. 

Results also show that there is a correlation between perceived persuasiveness of the Expert and the extroversion dimension of personality ($R^2=0.136$, p = 0.008). 
The positive correlation indicates that higher extroverted people are more likely to be persuaded by the Expert robot. However, no other correlation was found regarding other personality dimensions or regarding the perceived persuasiveness of the Joker. Although previous studies have found positive correlations between persuasive strategies and agreeableness as well as emotion stability \cite{anagnostopoulou2017exploring}, we could not confirm them in this study. This might be attributed to our limited sample size or due to the nature of the persuasion task.

Moreover, we hypothesized that personal characteristics might play a vital role in being persuaded by one specific type of power strategies, but this difference did not stand out in the results. A potential reason to this might be a number of hidden factors other than what we measured in this study, such as the need for cognition~\cite{cacioppo1982need}, which might have influenced the participants' decision-making. 
Need for cognition is a personality factor that taps individual differences in the tendency to enjoy thinking and to engage in abstract deliberation.
Since processing the expert argumentation requires higher level of thinking, people ranked higher on this factor might prefer the Expert.
Finally, we asked the participants to indicate to what extent they are willing to follow the suggestion of each robot in the future (FC).
The results evidenced that people were more eager to follow the Expert's suggestions even though no significant differences between persuasiveness of the two robots were found.
This could be due to the expertise of Expert and the fact that he stated more logical and rational statements. 
In other words, we can infer that people found Expert more reliable in the future context.
In spite of this, the Joker was perceived equally persuasive as Expert. Thus, we can conclude that the Joker's persuasion was based on the effect of the reward strategy, not the information. To put it another way, some people can be easily persuaded by means of rewards. 
Further, this finding also highlights the role of ``reward power strategy'' in persuading people: Although people found the Expert more trustworthy to be followed in the future,
the Joker was also equally successful in persuading them to choose its coffee. 

In addition, we would like to highlight that the obtained results do not depend on the CDH of the participants. Results of Chi-square tests revealed that no significant association exists between CDH and robot preference (LikingCoffee: $X^2(4)= 5.180, p= .269$; CoffeeTimes:  $X^2(4)=7.604, p=.107$) or coffee selection (LikingCoffee: $X^2(4)= 1.958, p= .743$; CoffeeTimes:  $X^2(4)=3.942, p=.414$). Moreover, no association exists between CDH and satisfaction or perceived social power of any robots. One might argue that people who like coffee might be more sensitive to its quality and would opt for coffee advertised from by the Expert, but the results did not confirm this. Another potential reason for this might be the Expert's arguments, which address the flavor. Hence, people who do not like coffee flavor might opt to go with one of the other two options instead. Hence, we suggest that future studies could consider other characteristics of coffee apart from taste and flavor.



Furthermore, we asked the participants to indicate on a 5-point Likert Scale how they perceived the robot having social power.
Regrettably, the question was hard to understand for most of the participants and the experimenter was asked a number of times about the meaning of ``social power.'' To be more specific, measuring social power of the robots might not be truly reflected using a single question. Thus, we skipped this question and excluded it from the analysis.

\subsection{Exploratory Findings}
In this subsection, apart from what was alluded to earlier~\cite{hashemian2019persuasive}, we investigate the effect of other factors that might influence the results and were not reported earlier.

As discussed earlier, evidence suggests that when there is a power match between the persuader and the persuadee, i.e., when both sides have high power or low power, there is a higher chance of gaining compliance~\cite{dubois2016dynamics}. Hence, apart from the social power question, we asked the participants to fill out the  personal sense of power (PSP) questionnaire~\cite{anderson2012personal}. 
Our results suggest that for the Expert, this difference was not statistically significant (t[49] = -.095, p = .925); this conclusion was derived based on power scores (power match exists: M = 3.62, S.E.= .224; no power match exists: M = 3.59, S.E. = .204).
Similarly, a higher mean score of persuasiveness was observed for the Joker when there was a match (M =  3.59, S.E. = .230) vs. no power match  (M = 3.23, S.E = .237).
However, this difference was not statistically significant (t[49] = -1.071, p = .290).
%

As social power of robots was not measured reliably in this study, we directly checked if PSP measure is associated with persuasion and found only weak correlation between PSP and the Joker's persuasiveness (r = .390, p = .005).
Further, to investigate if the personal sense of power leads to any effect on the perception of each robot's power, we checked potential correlation between these two factors.
The result does not lead to any correlation between PSP and social power (Expert: r = .111, p = .436; Joker: r = .123, p = .391). 

Additionally, we asked the participants to rate their satisfaction regarding the coffee they opted for. We expected to observe higher satisfaction in the participants with higher level of social power (based on the PSP scores). However, in contrast to previous evidence~\cite{mourali2013powerful}, we could not verify this relationship between satisfaction and personal power sense (r = .205, p = .149). We split the participants by the median score of PSP and compared the two groups regarding the reported satisfaction. We could not find any difference among the two groups. Although the average satisfaction was higher in the group of high-power people (Low power: M = 3.91, S.E. = .173; high power: M = 4.17, p = .172), the difference was not statistically different.

\subsection{Qualitative Analysis}
At the end of the questionnaire, we added an open-ended question asking the participants why they had selected the specific coffee. Answering this question was not mandatory, but only seven people skipped it. Qualitative analysis of the obtained contextual data provides new insight over the preceding results. 


The qualitative analysis was conducted to better understand the participants' motivation behind their decisions. 
We used a combination of the conventional and the summative approach proposed by~\cite{hsieh2005three} to investigate how people experienced the interaction with the robot and made decisions under different circumstances and conditions of the study.

First, regarding the people who selected the third coffee (i.e., the control coffee), most of them selected this option because of feeling empathetic toward the robots. Specifically, four out of eight indicated sympathy toward robots in their decision-making. For instance, ``If I had selected one robot, the other robot would get [sic] sad. I selected the middle coffee not to make any of them sad'' or ``I did not want to break their heart, so selected the other coffee.''
Additionally, two people selected the middle coffee based on their curiosity: ``The middle coffee seems mysterious to me, so I chose it because I was curious'' and ``I hoped robots would comment my selection regardless of my choice.'' 
Regarding the other two, one was compelled by the two robots and wanted both the joke and the well-advertised coffee and the other was non-compliant and stated, ``I don't like advertisement[,] and they were advertising their coffee. So, I selected the one that was not advertised.''
From these statements, we can infer that the selection of this group was not based on a random selection, but rather, this coffee was selected due to equal persuasiveness of the two strategies (except for the non-compliant and the curious user). 

Among 22 people who selected the Joker's coffee, 20 people answered the open-ended question. Overall, only one participant mentioned that he was not interested in the coffee as he had a cup of coffee right before the experiment, so he selected the Joker. Moreover, 11 people indicated, in short, that they wanted to hear the joke. The rest provided more information and stated that they selected this option because of the Joker's social behavior and characteristics, such as its personality and sense of humor or its joy and emotional interaction. Two participants provided very interesting information: 
``Just because of an emotional decision instead of applying my rational mind'';
``I wanted to hear the coded joke. If the two would be [sic] humans[,] I would definitely not have chosen Emys (Joker).'' These statements highlight the role of Joker's strategy in persuading people for social robots.
In sum, owing to either the joke or its funny attitude (as an instance of social reward), the Joker could successfully influence a number of participants and manipulate them to select its coffee.

Finally, among the 21 people who selected the Expert's coffee, 12 users answered the open-ended question. These people can be categorized in two groups based on their answers: the ones focused on the coffee characteristics (such as good or bad, origin, roasted or not, ingredients) 
and the ones focused on the robot's behavior (highlighting facts, being knowledgeable, and displaying seriousness). For instance, the participants made following comments:
``He described it very well,''
``Emys expressed why he thought his coffee was better,''
and, more interestingly, ``Emys looked like an expert.''
Altogether, these statements prove that the Expert could persuade the users by using its expert social power strategy and influenced them to select its option among the others.

Furthermore, we asked the people who selected the Joker to rate the joke out of 5. 
Two people did not find the joke funny; however, this did not affect their satisfaction negatively (one was moderately satisfied [4] and the other was somehow satisfied [3]). In fact, one participant who found the joke a bit funny reported the least score for its satisfaction (2 out of 5). He is the only participant whom we can suspect to be unsatisfied due to receiving an unfunny joke.

Finally, we would like to highlight that there was not always an agreement between the robot that the users followed and  the robot that they preferred (refer to Table~\ref{table:st1_other} for more details). Some people followed the advice of one robot but preferred to interact with the other one: ten people who selected the Expert's coffee indicated higher preference toward interacting with the Joker, and six people stated they preferred interacting with the Expert but selected the Joker's coffee. Only 25 people agreed that they preferred to interact with the robot promoting the coffee they selected.

Regarding the former group, it is not unlikely that the participants selected the Expert's coffee as they found it a better choice but preferred to interact with the joker as it was more friendly and funny. Thus, they were persuaded objectively by the Expert's expertise and compelled subjectively by the Joker's social reward (funny and lively interaction).
The latter group selected the social reward and focused on hearing a joke but claimed that they preferred the Expert. These people were persuaded objectively by receiving the joke as the social reward but were attracted subjectively by the expertise of the Expert.
These findings support that the two robots were persuasive, either in one way (subjectively or objectively) or both ways.

\begin{table}[!t] 
\caption{Study 1: Robot Selection vs. Robot Preference}
\label{table:st1_other} 
\centering 
\begin{tabular}{llr}
\toprule
Selected Robot & Preferred Robot & Count\\
\cmidrule(r){1-3}
	    &Expert	&9\\
Expert  &Joker	&10\\
	    &Neither	&2\\
\cmidrule(r){1-3}
	    &Joker	&15\\
Joker   &Expert	&6\\
	    &Neither	&1\\
\cmidrule(r){1-3}
    	&Neither	&1\\
Neither &Expert	&3\\
	    &Joker	&4\\
\bottomrule
\end{tabular}
\end{table}


\subsection{Summary of Findings}
In summary, in Study 1, we investigated the influence of two different persuasive strategies in an adversarial setting. To do so, we performed a user study regarding an actual decision-making process within a persuasive setting. Our main goal was to examine the effect of different persuasive strategies that are based on social power. The second purpose of the study was to investigate the perception of people having different personalities regarding such persuasive strategies. To the best of our knowledge, the use of social power as a persuasive strategy has not been explored before this study. 

Altogether, the results of this study provide important insights into persuasion in HRI.
First, this study identified two different persuasive strategies that were selected and preferred equally. However, these strategies led to different perception of robots based on personal characteristics of each user, such as their personality also affects which strategies are deemed to be more effective. 
 
The second major finding was that using social reward is effective. To be more specific, in the two persuasive settings, the user was rewarded ultimately by receiving a coffee capsule, whether they selected any of the two promoted coffees or none of them. However, selecting the Joker's coffee yielded another dimension of reward, hearing the joke, as an example of a social reward.  
Undoubtedly, social rewards are cheaper than monetary ones and are easily applicable in any type of social robots. The result of this study shows not only its effectiveness but also it applicability in persuasion.

These findings suggest that, in general, robots are capable of persuading people; however, personal differences should be taken into account. Further, it should be noted that only two bases of power have been tested here, and the rest are to be examined in future attempts. The result of the current study showed that the two strategies used were preferred equally; however, it should be noted that different power strategies might lead to different outcome. Also, the level of power exerted might influence the results. For example, a stronger reward strategy might be preferred more. In other words, the comparability of such power strategies is inherently problematic since the power of an implemented strategy depends to a large extent on its concrete implementation.

In sum, the key contributions of this study are 1) testing the persuasive strategy effectiveness in a real-choice task and 2) having a within-subject design that allows for testing competitive persuasion. 
This study tested the effect of persuasion in an incentivized real-choice task, which increases the external validity of the design and has an implication for robotic persuasion in a consumer choice setting.
In the task, the participants chose a coffee capsule after interacting with the persuasive robots, which is an advancement over hypothetical choices used in other research studies on robotic persuasiveness. In addition to the real choice, the study also measured participants’ willingness to follow the robots’ advice in the future, which can potentially reveal a difference in short-term and long-term persuasive results.
Moreover, the study examined the participants' perception of the robots' warmth and competence, which offered opportunities to understand the mechanism of how social power strategies affect persuasion. 

In addition, the study used a within-subject design in which two robots adopting two different strategies interact with the human subject at the same time. Given the sample size, the within-subject setting increased statistical power for comparisons between the strategies. Further, this design also provided a unique opportunity to test the effectiveness of the persuasive strategies in a competitive persuasive setting.



\subsection{Limitation and Lessons Learned for Designing a Future Study}
After analyzing the overall results, we acknowledge a number of limitations of the study, which can be helpful in our future research.
First, the specific design of this study only allowed for comparison between the relative effectiveness of the two strategies. With the current design, we cannot compare the two strategies, and the two may work together for the observed effect.
It might also be fruitful to conduct a study to determine the effectiveness of each strategy and to directly measure the persuadability of the robots. 

Second, we attempted to measure social power of the participants using the task-specific questionnaire, but no correlation was found in the collected data regarding this single question. We argue that information about power level of each side would give us a better understanding of the interaction. 
In general, the individuals have been categorized under two psychological states: high power vs. low power. On the other hand, individuals dealing a negotiation were assigned two roles: communicators (those who deliver message) vs. audience (those who receive a message, or the targets)~\cite{dubois2016dynamics}. The power level of each side, either the communicator or the audience, affects the result of the persuasion attempt. Thus, it is necessary to measure social power levels more profoundly than what we did here in order to provide evidence supporting that participants’ perception of the social power of the robots increased because of the two strategies.

In other words, manipulating humor and expertise might not warrant the achievement of social power. We need further investigation to better comprehend how the participants perceived the robots, for instance, if the expertise gave expert social power to the robot. As stated earlier, to measure the power level of each robot, we used a single item in the questionnaire. However, measuring social power of the robots in this way might not lead to reliable findings. One potential reason for this might be misunderstanding in the interpretation of ``social power'' expression; this interpretation must be same for all the participants.

Similarly, we need to verify if the joke gave reward power to the robot. Hence, in the future step, we would like to apply a standard questionnaire of social power. Specifically, a future study is required to investigate if telling jokes counts as a social reward. 
To be more specific, telling a joke might promote liking toward the source of humor and hence induce referent power. Thus, it must be considered carefully using self-reports to determine if the participants perceived the joke as a reward.

As mentioned earlier, as the control condition we put a third coffee to decrease the probability of selecting randomly one of the two strategies. Simply put, having three options decreases the randomness probability to 33\% (compared to 50\% in case of two options).
A more suitable control condition could have been ``absence of power strategy'' or ``neutral product presentation''; however, in the specific design, it might have led to a silent robot or a less intelligent robot, leading to a bias toward the other robot.
Hence, we introduced a condition that excludes the presence of both the robot and the strategy. As a consequence, the responses observed in the control condition may not be interpreted as a result of the absence of power strategy only. 
These limitations motivated us to design another scenario in which only one robot interacts with people.

\subsection{Recommendations for Further Research}
Apart from these modifications, we suggest other avenues for potential future research.
A potential question raised by this study might be that if combining the two strategies would lead to higher persuasion, it is worth investigating in the future. 

%
Recent studies have found correlation between ostensible gender of the robot and perceived persuasiveness~\cite{ghazali2018effects, siegel2009persuasive}. Although EMYS does not clearly appear to be either female or male, the two voices we assigned to the robots were both male voices. A potential future work worth performing is using voices with different genders to see whether its combination with persuasive strategies leads to a higher effect.
Moreover, in this study we did not measure the trust toward the robots. Investigating the potential interrelation between trust and persuasiveness of the robot would be of great value. 
%

Further, when people are subjected to strong persuasive attempts, they may respond negatively toward the attempt, a behavior that is known as psychological reactance~\cite{ghazali2018influence}. 
A future study could assess this by measuring the strength of the perceived persuasiveness message of the robot from the perspective of the participants. Also, participants' culture and background may affect how they perceive the over-the-top language used by the expert. 
Thus, a further study could also assess the effect of subjects' trust regarding such arguments~\cite{hashemian2019persuasive}. 


One important question regarding these kinds of persuasive social interactions with robots is related to the long-term effects of several persuasive attempts, which might be a fruitful avenue for further research. In addition, future studies should be performed with a more homogeneous (gender-balanced) sample. Further research should focus on using more specific questions about the perception of the joke and if the subjects find it as a positive reward or if they really find the other robot as an expert. Furthermore, a social power scale is required to implicitly measure the perceived level of social power, or validating the dialogues by using experts'/judges' criteria may resolve this issue. 

We videotaped the sessions using two cameras to record the participants' behavioral and non-verbal responses apart from the self-report measures (e.g., perceived persuasiveness) in the final discussion of the results. This would be a fruitful area for further work.
Further research could usefully explore the participants' social responses toward robots' persuasive messages, using participants' behavioral cues and body language, facial expressions, gesture and postures, to further investigate their decision-making process while facing the two power strategies. 
A more balanced discussion could be achieved by giving more importance to the behavioral results (i.e., the actual decisions that participants made) and by considering the self-report measures, just as a source of hypothesis to be tested in future studies.

In a nutshell, according to French and Raven's theory, power arises from different sources. In this study, we equipped robots with two different sources, i.e., reward and expertise, and designed them so as to generate persuasive strategies based on their power sources.
Overall, this study shows that using different sources of power, and hence power strategies, appears to be an equally viable solution to designing social robots capable of persuading people. 
Moreover, we argue that the result of this study signify that social rewards can be effective at persuading users and that unlike material rewards, they are unlimited and always available.

%
Although the rest of power bases similarly lead to corresponding persuasive strategies, they are left as future work. The two strategies were selected mainly because they were most applicable in the context of our designed task. 

\section{Study 2}\label{study2}
The previous study proved that social power can be used as a persuasive strategy for social robots. However, with the specific design of the study, we could only infer which robot is preferred over the other one. 
To be more specific, Study 1 compared the effectiveness of the two forms of social influencing strategies (rewards vs. expertise); in addition, it would be interesting to (separately) show the effectiveness of each of the two.

Additionally, as we discussed earlier, despite the acknowledged role of message strategies in persuasion, little is known about how social robots' attempts may achieve higher persuasion using such strategies. 
Earlier studies on HCI examine compliance gaining behavior (CGB) in interpersonal persuasion.
Evidence shows that four strategies of emotion, logic, reward, and punishment are effective in persuading computer-mediated communication (CMC)~\cite{wilson2003perceived}.
Further, in the field of HRI, previous research has established that two of these strategies---emotion and logic---lead to higher persuasion~\cite{saunderson2019would}. However, less is known about the reward and punishment strategies. 

\begin{figure}[!t]
\centering
\includegraphics[width=0.9\linewidth]{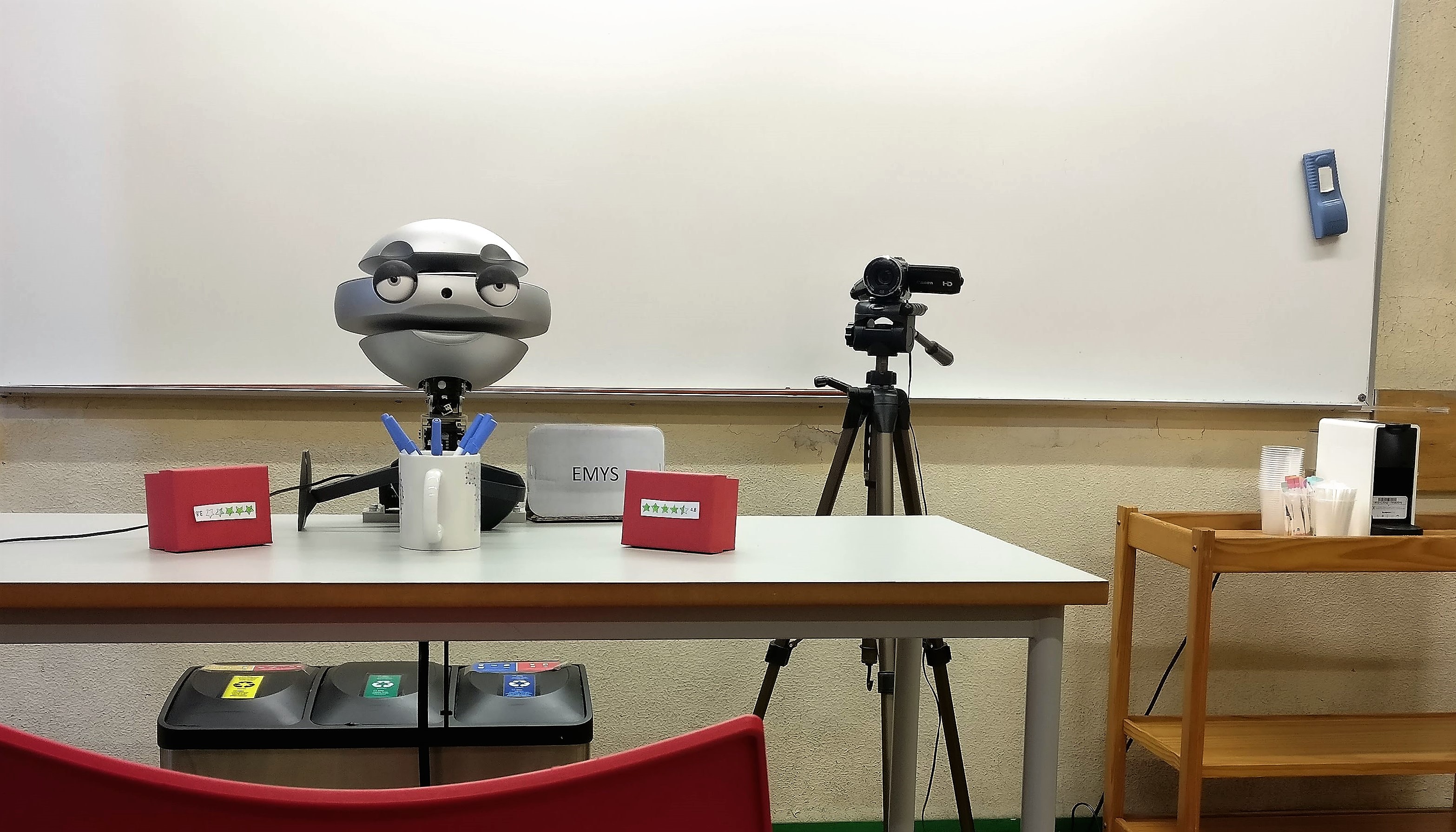}
\caption{Study 2: Setup.}
\label{fig:st2_setup}
\end{figure}

Hence, we designed another study to further investigate persuasiveness of social robots using social power strategies. 
In this design, a single robot attempts to persuade the users in two different conditions, which are compared to a control condition (three conditions in total). In one condition, the robot aims at persuading the users by giving them a reward; in another, it tries to persuade them by punishing. In the control condition, the robot does not use any strategy. (Fig.~\ref{fig:st2_setup} depicts the setup of the second study.)

Specifically, in Study 2, the robot presents two coffee capsules hidden in two boxes and labeled with the star classification based on reviews of other people. In the control condition, the robot asks the participant to select any of the two coffees that they prefer. In the reward condition, it offers a pen to the participant if they select the lower-rated coffee. In the coercion condition, it first gifts the participant a pen and then asks them to return it if they select the higher-ranked coffee.

In Study 2, we would like to investigate how different levels of reward and punishment may affect the persuasion. To investigate the effect of \textit{loss} on the persuasiveness of the robot, we consider two different coffee ratings. In one scenario, we assign a rating of 3.8* vs. 4.8*, and to resemble a higher loss, we assign a rating of 3* vs. 4.8*. To be more specific, selecting a 3* coffee has a higher probability of receiving a bad coffee, i.e., a loss in achieving a better coffee.
Overall, considering two different levels and three conditions, we designed a 2$\times$3 between-subjects study to investigate these effects (we call each sub-group of the study as follows: 3*/3.8* $\times$C/R/ctrl).

As discussed, we explore the potential effect of the two other strategies, i.e., reward and punishment (coercion), in a persuasion task. This section discusses the empirical study we conducted with the goal of understanding the extent to which these strategies used by social robots are persuasive in influencing a person's choice while facing a better vs. a worse option. Thus, in this design, we investigate the effect of message strategy on participants' decision-making while they face two comparable options in an interpersonal persuasion with a robot. 

Here, we investigate the following hypotheses:
\begin{enumerate}
    \item \textbf{H1}: A stronger punishment would lead to a lower compliance. In other words, under the same circumstances, when a robot's request leads to a higher loss, we expect the human user to be less compliant to the robot, facing a higher loss due to the punishment.
    \item \textbf{H2}: Coercive strategy leads to higher persuasion. Here, inspired by~\cite{kahneman1991anomalies}, we hypothesize that people would be more sensitive to losing an owned reward than gaining a reward.
    \item \textbf{H3}: Coercive strategy decreases warmth and increases discomfort. We expect the participants to perceive the coercing robot negatively as it imposes a penalty~\cite{xiao2005emotion}.
\end{enumerate}

In the remainder of this section, we present a summary of the results of the user study performed to investigate if social robots are able to persuade people to opt for a less favorable choice (for more details, see ~\cite{hashemian2020roman}). We initially compared two different conditions (persuasions) with a control group and then investigated the difference between the two strategies by comparing the persuasion groups. 

\subsection{Results}\label{st2-results}
At the end of the experiment, 90 people (38, or 42.2\%, females and 52, or 57.8\%, males) participated in the experiment voluntarily. To start with, we first checked if the \textit{difference in coffee ratings} would influence the decision-making of subjects. 
Specifically, we hypothesized that the higher difference in ratings would lead to lower compliance, i.e., a higher difference between the scores would lead to a higher risk of receiving a bad coffee (lower-ranked coffee). We assumed that the high difference in the ranking presents a higher risk, thereby leading to a higher resistance to the persuasion. Since the level of reward/coercion is fixed but that of the resistance is not, it hence leads to lower effects of reward and coercion in conditions with lower rankings. We presumed that this effect leads to less compliance. 

However, the results of logistic regression tests implied that coffee rankings used in Study 2 are not a good predictor of decision-making of the participants. This means, although the robot could persuade a large number of people to select the lower-ranked coffee (57.8\%), this difference was not significantly higher in the 3* vs. 3.8* ranking (Wald[1] = 1.255, p = .263). Hence, we reject the hypothesis H1. 
A potential reason for this finding might be the minor differences between the two rankings (3 vs. 3.8). A further study could assess this effect to determine a threshold so that the rating of the less desirable choice is not too  low or too  high; this would make decision-making easy. In other words, when the lower rank is too low, the participant might not risk and reject the persuasion easily. Conversely, when the lower rank is too high, it becomes very close to the other option, and the participant would accept the persuasion to benefit from the two rewards (a pen and a good coffee).

Furthermore, we hypothesized that coercion would lead to high persuasion compared to the control condition: 
On comparing the coercion strategy with control condition, the results indicated that the model is statistically significant (Wald[1] = 4.95, p = .026). In this case, extroversion (Wald[1] = 4.786, p = .029) and openness (Wald[1] = 4.330, p = .037) are good predictors of participants' choice to make decision.

On the contrary, the results indicated that reward is not a significant predictor of persuasion (Wald[1] = .029, p = .864).  
A potential reason might be the uncontrolled distribution of the participants in different conditions of the study.
Actually, a closer look of the data highlighted that people who had interacted with robots prior to this study acted differently compared to the others who were new to robots (novelty effect). Specifically, the results of a t-test showed that people who had already interacted with robots (M = 1.6, SD = .49) were more compliant and got persuaded by the robot (M = 1.33, SD = .47). 

Moreover, the collected data revealed that most of the participants who had already interacted with robots fell in the reward group; hence, the interaction effect of this earlier encounter with robots might have diminished the effect of reward. 
Furthermore, the gain of coffee itself on participating in the experiment already put the participants in somewhat of a reward situation, which could additionally have interfered with the actual reward strategy of gifting a pen. In other words, the participant will always get a free coffee (which they would not have had otherwise) but only get a pen if they take the lower-ranked coffee. 
An important factor here would be how much the participant values any of these two gifts, which unfortunately was not measured in this study.

%
We also hypothesized that coercive strategy would be perceived negatively (higher on discomfort), while reward strategy would be perceived more positively (higher on warmth). We assumed that giving a reward would make the robot more friendly, and in contrast, coercion would be a negative predictor of liking.
However, we could not verify this hypothesis based on the collected data, and the results indicated contrasting findings, as the coercing robot was scored higher on warmth. 
A potential reason for this might be that some of the participants did not perceive the coercive action of the robot as a punishment but rather perceived it as being funny and laughed out loud after the robot asked them to return the pen. Another potential reason behind this might be that the coercive action was weak, since the participants had no intrinsic attachment to the pen. Another factor might be the minor differences in the dialogues. 
Particularly, we hypothesized that different strategies will affect the participants differently, hence causing a different perception of the robot.
However, actually, the two scenarios differed only in two sentences. Further, in the two scenarios, the robot showed the same instances of social interaction, such as facial expressions and gaze.
Conversely, to get perceived negatively, the robot needs to show samples of a bad attitude, for instance, being rude. 
Hence, we cannot confirm the hypothesis. 
However, this finding should be interpreted cautiously. To be more specific, we could not verify all presumptions of the test due to the bias of earlier interaction. Hence, the results might not be generalizable to other studies. 
Hence, this hypothesis needs to be further investigated in different scenarios with significantly different dialogues and social cues (and probably with longer duration of time).

\begin{figure}[!t]
\centering
\includegraphics[width=0.9\linewidth]{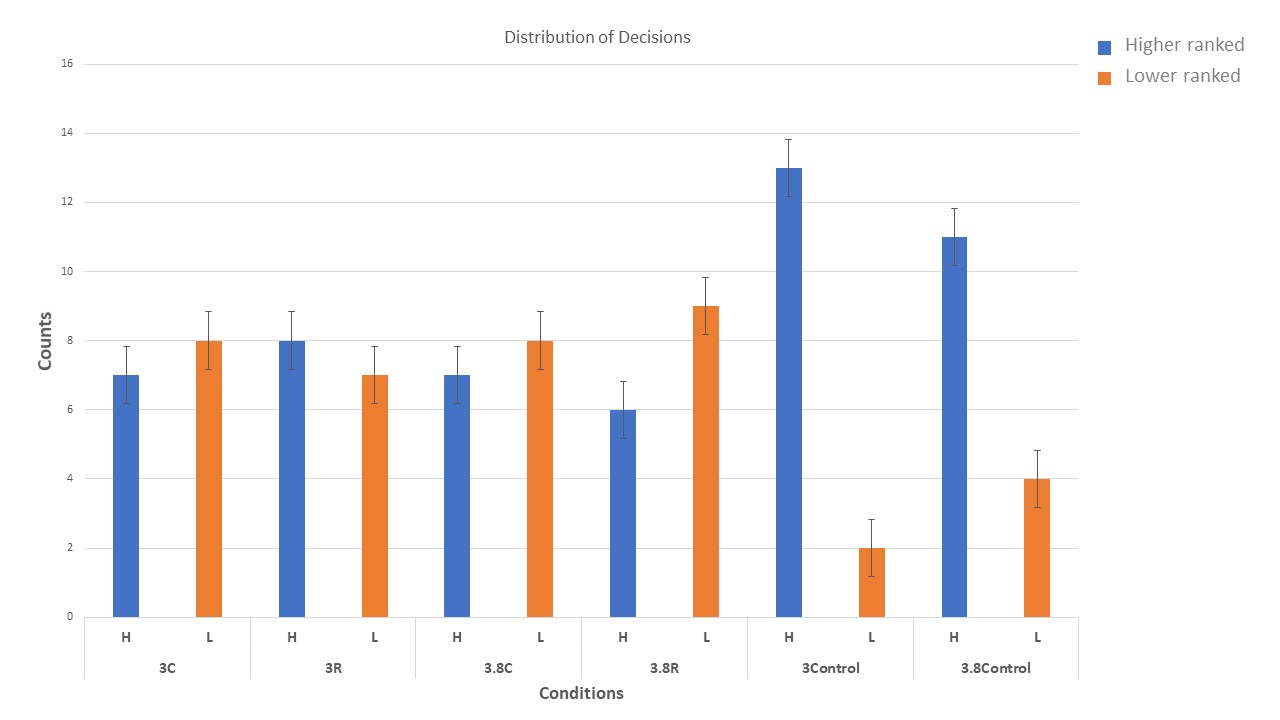}
\caption{Study 2: Distribution of coffee selection over the six groups.}
\label{fig:dec_dist}
\end{figure}

Further, in Study 2, we measured persuasion both objectively (the selected coffee) and subjectively (robot perception). The results indicated that the reward and punishment strategies make the robot more persuasive, as measured objectively through user compliance with the robot's request (objective behavior). As depicted in Figure~\ref{fig:dec_dist}, the lower-ranked coffee was selected less frequently in the two control conditions. 
However, we could not verify if these strategies impact participants' perception of the robot's persuasiveness and social attributes (subjective perception).
In sum, although the robot could objectively persuade the users (to select the lower-ranked coffee), the subjective facet of persuasion was not significant. 
A potential reason could be the difficulty in accurately measuring the perception of a robot using subjective measures~\cite{winkle2019effective}.

%
Additionally, recent evidence suggests that one single factor may have different influences on persuasion: In one circumstance, it might influence the degree of elaboration; in another, it might influence the valence of elaboration, while in the third situation, it might serve as a peripheral cue~\cite{petty1986elaboration}. These differences can give rise to different effects on persuasion and hence inconsistencies in research finding considering a single factor. Hence, further investigation is required to determine in which direction the persuasion influences the user. 

Thus, we expected that the difference in ratings would lead to different levels of power, thereby leading to higher persuasion in case of higher loss. 
However, the results of this study failed to verify this hypothesis, so we designed the next experiment with a significant valuable reward, as described in the next subsection.




\subsection{Exploratory Findings}
In this subsection, apart from the previous hypotheses~\cite{hashemian2020roman}, we further investigate the effect of other factors that might influence the results. 
To this end, we investigate the role of the participants' cultural background and its relation to the language used by the robot. 
In doing so, we consider the nationality of participants as a general measure of culture. 
In this study, the participants were from ten different ethnicity: Portuguese (70\%), Iranian (15\%), and the rest (Angolan, Brazilian, Chinese, French, German, Guinea Bissau, Ukrainian, and American). 

Based on a seminal work by anthropologist Edward Hall~\cite{hall1959silent}, cultures (as well as language constructs) are generally categorized into ``high-context'' and ``low-context'' cultures. As per this perspective, different cultures lie on a continuum based on how explicit messages are exchange as well as how much the context is important in communication. In other words, in some cultures, a message is delivered mainly using words, whereas in others, the context and the way a message is delivered (using non-verbal cues) also affect the meaning of sentences. (For instance, in Asia or the middle-east, messages could be delivered through more indirect ways.)

A non-small body of research in HRI investigates whether the conclusions found to be true in human--human interaction contexts are also valid in HRI scenarios when the robot is acting in accordance with the rules of a culture. As an example, it has been found that the wording of a sentence (explicit/implicit) impacts people differently according to their culture and leads to different results in terms of how likely a person is to follow the robot’s advice~\cite{bruno2017paving}. 

As the study was performed with mostly non-native English speakers, we verified that there were no statistically significant differences in the level of English proficiency among the six groups (F[5,89] = 1.013, p = .415).
Moreover, we checked if the nationality of the participants (as an estimate of their culture) had no effect on the results (F[9,89] = .810, p = .609). 
Further, as we had few samples from some ethnicity, we divided the participants based on high- and low-context cultures~\cite{hall1959silent}. The result of the analysis of covariance (ANCOVA) indicated no significant differences among people of high- and low-context culture regarding their decision-making (F[2,89] = .401, p = .528).

Finally, we checked if the personality traits of the participants influenced their decision-making and added them as covariates in the analysis. The results did not show a significant difference in decision-making due to personality traits, but extroversion demonstrated a marginally significant effect (F[1,89] = 3.409, p = .069).

In addition to cultural differences, according to what was discussed earlier, as the study was inspired by social power theory, we measured other factors such as robot's social power and PSP to better understand how participants perceived the persuasion and if the persuasion evoked any social power in the robot. However, these factors were not predictors of decision-making, nor significantly affected the perception of the robot (PSP: F[1,85] = .732, p = .395; social power: reward: F[1,84] = .556, p = .458; coercion: F[1.84] = .149, p = .701). 
%
Hence, to check to what extent the participants find the robot rewarding/coercing, a more direct questionnaire would be required in a future study.

%
We devised a number of questions in the questionnaire designed specifically for this task to further investigate the perception of the participants. 
Specifically, we asked the participants to answer the following questions on a 5-point Likert scale: 

\begin{enumerate}
    \item How persuasive did you think EMYS was?  
    \item Consider a situation in which you have an opinion different from EMYS's; will you change your opinion in such a way so as to be consistent with EMYS's? 
    \item Imagine a situation where EMYS gives you a bit of advice in the future. Please specify the likelihood that you would follow EMYS's advice?
\end{enumerate}

We averaged the scores against these questions and compared them considering the effect of earlier interaction with robots. The results suggested that the average scores against the task-specific questions were not significantly different across the scenarios (F[2,86] = 2.724, p = .071, $\eta^2_{p}$ = .060). They led neither to any significant or strong correlation with robot perception (W: r = .276, p = .009; C: r = .216, p = .042; D: r = .183, p = .086) nor to the final decision of the users (r = .062, n = 87, p = .567).

\subsection{Qualitative Analysis}
We also checked the answers to the open-ended question qualitatively, which are reported in this section.
Among the 90 participants, 73 answered the open-ended question; among these, 29 selected the lower-ranked coffee. 
Overall, six people were assigned to the control condition (two people in 3ctrl, and four people in 3.8ctrl). Most of these people indicated curiosity toward the rankings (e.g., ``To see if other people's assessment was correct'') or inherent non-compliance (e.g., ``I have free will to choose. It's just my rebel way of living'').

Moreover, 14 people belonged to the reward conditions (five subjects in 3R and nine subjects in 3.8R).
These people mostly highlighted the role of the pen or the gift as their main motivation to select the lower-ranked coffee. For instance, ``3.8 is not a bad rating[,] and the reward of a pen seemed worth the lower[-]rated coffee''. 
Interestingly, people in 3.8R highlighted the minimal difference between the two ratings and their interest in the pen. 
For instance, ``I don't care much about the coffee taste, since all taste more or less the same. Even if one is slightly WORSE for me, since I would get a pen[,] I would prefer the WORSE one.''
In sum, they selected this option as they found it ``more rewarding,'' since they received coffee as well as a pen. 

Finally, nine people in the coercion condition (four people in 3C and five people in 3.8C) indicated an attachment and interest toward the pen. For instance, ``It was a good coffee despite being the lower[-]ranked one[,] and I could keep my pen'' or ``3.8 is not a bad score[,] and I get a pen''. 

The remaining 44 people selected the higher-ranked coffee; among these, 24 belonged to the control condition (13 in 3ctrl and 11 in 3.8ctrl).
Most of these participants highlighted the higher rank of the coffee as their motivation. Although some suspected the credibility of the rankings (``If it's ranked higher, the chances of being better are higher, although this depends on how many people rated it''), they still did not want to risk receiving the bad coffee (``Higher probability of being good because of rating'' or ``There is no reason to pick the WORSE[-]rated one. Even if I don't trust the robot[,] picking left [higher-ranked coffee] is not WORSE than a blind pick.'')

In addition, nine people in the two reward conditions (four in 3R and five in 3.8R) selected either based on their curiosity (``Just to see why it had such a high ranking'') or because they were simply not interested in the reward (``Do not need the pen''). Moreover, they valued the coffee more than the pen (``Because, assuming the ratings are correct, I prefer having a better coffee than a [bad] coffee and a pen'').

Finally, 11 people were assigned to the coercion condition (four in 3C and seven in 3.8C).
Most of these people highlighted the low value of pen (e.g., ``A pen is not worth drinking bad coffee'' or ``I don't need a new pen, and prefer better coffee''). 
In addition, one subject was curious about the high rank of the coffee (``To know if the rank it was correct or not'' [sic]).




%

\subsection{Summary of Findings}
%
As discussed in~\cite{hashemian2020roman}, the data analysis highlighted a significant difference (t[88] = 2.469 , p = .015) between coffee selection (if the subjects selected the higher- or the lower-ranked coffee) of participants who had already interacted with any robot (M = 1.6, SD = .49) vs. the others (M = 1.33, SD = .47). People who had already interacted with robots were more compliant and got persuaded more by the robot. 
To overcome this effect and the potential bias of prior interaction with robots, we considered this confounding variable as the covariate and included it in one-way ANCOVA (in case of continuous dependent variable, i.e., RoSAS questionnaire) and logistic regression (in case of categorical dependent variable, i.e., participants' decision-making or which coffee they selected).
It should be noted that there was no significant difference regarding prior interaction with EMYS robots (t[88] = 1.54, p =.128).

Withal, the results showed that the robot has the potential to persuade the users and make a bias on their decision making. To be more specific, comparing to the control group, in which no persuasion was used, the robot could bias a number of participant's decisions toward a less-desirable choice (Wald(1)=6.627, p=.010). So, the robot could change people's behavior in the expected direction. 

However, the subjective measures used in this study did not yield significant findings in our expected direction and that would be a fruitful area for further work. To measure the subjective perception of the participants, we applied the RoSAS questionnaire~\cite{carpinella2017robotic}. We postulated that coercive strategy decreases warmth and increases discomfort. the findings failed to verify this hypothesis. To be more specific, regarding the discomfort scores the results did not indicate any significant difference between the score of discomfort in any conditions (p=0.543, effect size: 0.053).

Additionally, the results of different ANCOVA tests show that 
there is a statistically significant difference in the scores of warmth between the coercion scenario and the control (p=0.039, effect size=0.227). And, surprisingly, this score is higher in the coercive condition (Coercive condition: M=4.41, S.E.=.17; Control condition: M=3.91,S.E.=.33) which is in contrast to our expectations.

Furthermore, the results of this study did not find any evidence that a stronger loss would lead to higher persuasion (Wald(1)=.266,p=.606). However, the findings indicated that coercion is a good predictor of decision making (Wald(1)=5.692,p=.017).
%
%

\subsection{Limitation and Lessons Learned for Designing a Future Study}
One source of weakness in this study is that the results failed to indicate how the users perceived the robots in terms of social power. We require more evidence that the robot's social power is manipulated. In other words, we require a specific questionnaire measuring this more carefully. In addition, the design of task-specific questions did not establish any significant finding and should be verified using more attentive questions. 

Furthermore, a stronger manipulation check would be of great value to see how the participant perceived the pen as a reward/coercion and how much they value each of them  (as seen in the qualitative analysis). Measuring how much participants were attached to the pen and how much they desired having coffee would enhance our understanding of their behavior.
In other words, as discussed earlier, the coffee itself was a gift in the experiment. 
Depending on how much participants actually like/wanted the coffee might affect their assessment of the options presented by the robot.

We have made an attempt to examine both the role of the persuasion actor (social robots) and the persuasion target (human participants) in its theoretical model by measuring the personality of the subjects. However, the result of this study did not yield any findings. 
This might have happened due to the small number of participants in each bin. Collecting a higher number of data might open up more insight in this direction. Also, due to the interaction effect of previous interaction with robots, we had to apply a logistic regression to analyze the data. This test also requires a large number of samples. 

Another limitation of the study might be the design of the control condition. In the current design, in the control condition, the robot lets the subjects select an option freely without exerting any power. This gives us a baseline of decision making in the absence of any power or persuasion. Another control condition could be designed in such a way that the robot asks the participants to take a worse choice with no persuasive strategy. Rather than letting them to freely select their coffee.

Finally, to have a more interactive and believable scenario, we designed dialogues dependent on the participants' responses. For instance, in the beginning of the interaction, the robot asks the subject if s/he has already met the robot. Depending on the answer of the subject, the robot responds differently to induce the illusion of having a real-world interaction.
Specifically, if the participant has seen the robot before, the robot responds with a personal affective statement ``I am very pleased to meet you again''. This could be perceived to show goodwill, shown to influence robot persuasiveness~\cite{winkle2019effective}.
Also, when the subject does not provide any response when questioned,  the robot says ``I didn't hear you''. However, depending on the specific interactions and what triggered this behavior, this could reduce robot credibility by suggesting a technical error/lack of understanding compared to the positive/negative responses.
Finally, the reward dialogue was implemented with an additional affective signal (joy animation) with no equivalent present in the coercion strategy. These minor differences might have influenced the perception of the user and might have affected the results.
In sum, under specific dialogues, some of the condition/response-dependent dialogue may have additionally influenced persuasiveness and caused inconsistently with the main reward/punishment strategies. While, similar displays are not present for the other response conditions. 
This might have indirectly biased the participants perception and responses that was not aimed by this study (full dialogues are listed in Table~\ref{table:st2_utterances_apdx} in the appendix).

As future work, the study could be repeated using a higher number of participants who already interacted with robots. Or it might be applied to people new to robots in multiple sessions to decrease the novelty effect. 

Considering the current dataset, we are not sure how people perceived the strategies and further work needs to be done to establish this. Also, we have recorded the behavioral and non-verbal responses of the participants using two cameras. Behavioral analysis of the user would be of great help in determining their perception. It would be interesting to see if people's susceptibility to persuasion, specifically coercion, or to reward, would have an impact.

This design could be extended to other studies.
For instance, the current design might provide an opportunity to investigate the ``endowment effect'' and ``loss aversion''~\cite{kahneman1991anomalies} theory in a future study. 

More broadly, research is also needed to determine a prior validation of the dialogues to check if they lead to the desired power sources. Moreover, the task-specific questions were designed in a direct way and might influence participants to respond by social desirability. 
Finally, since the robot does not physically interact with the participants, it might be a good idea to compare the results with a virtual character or in a control condition without any robot.

\section{Study 3}\label{study3}



Having discussed the necessity of a new study, this section discusses the design we used to address a number of limitations of the previous studies~\cite{hashemian2021persuasive}. 
In this study, we only focus on one of the bases of power, i.e., ``reward base'' which is in common with the two previous studies. Other bases could be investigated in a similar approach in future studies. 



\subsection{Design}
This design is inspired by a conceptualization of power introduced in~\cite{hashemian2018enhancing}.
Based on Equation 1 in~\cite{hashemian2018enhancing}, this model indicates that reward power has a linear relationship with the amount of promised reward (\textit{rew}), probability of giving the reward (\textit{p}) and the the way the actor induces (\textit{induction}) the rewarding action (equation~\ref{eq_reward_}). 

\begin{equation}\label{eq_reward_}
        Power_{rew} = rew \times p \times induction 
\end{equation}

Hence, having other parameters fixed, increasing the value of the reward increases the force of social power. Also, considering a proportional linear relationship between social power and persuasion, this increase in power leads to higher persuasion (to some extent before a reaction happens).

The main research questions of this study are 1) to analyze how the different levels of reward influences the decision making of participants and 2) how this effect changes over a series repeated interactions. 3) if the novelty effect would influence the decision making in this design.
To answer these questions, we devised a mixed-design study within a decision-making scenario, in which we manipulated the level of rewards a robot gives to participants. To be more specific, the study contains two reward values (levels) and two control conditions: one with zero reward and one with no interaction with the robot (one-fourth of the participants were assigned to each group). In other words, in one control condition, social power is not activated; and in the other control condition, social power is activated but without the presence of a robotic persuader. 

In this design, after a decision making process, the robot tries to persuade the user to change their mind and select another alternative. 
To persuade, the robot uses a reward social power strategy and the task is repeated to investigate if the effect of social power on persuasion decays. 

In sum, considering formula~\ref{eq_reward_}, in the designed experiment, we assume that \textit{p} and \textit{induction} are fixed (as explained later).
And we manipulated two independent variables: one is the reward the participants receive. And the other variable is presence/absence of a social robot. 
We also considered two dependent variables: 1) the participants' decisions or if the they accept/reject the offer (objective measure), 2) how the participants perceive the robot (subjective measure).

\subsection{Hypothesis}
In this context, we expect to observe the following outcomes:

\begin{enumerate}
    \item \textbf{H1.} Higher social power (resulted from higher social reward) leads to higher persuasion.
    \item \textbf{H2.} People who are new to robots might be affected by the novelty effect. And this effect might interfere with the manipulation and diminish the effect of higher social power utilized to persuade.
    \item \textbf{H3.} Over a repeated interaction, the effect of power on persuasion does not decay, considering that the level of power is fixed.
    \item \textbf{H4.} Giving rewards increases the robot's likeability.
    \item \textbf{H5.} The presence of robot leads to a higher persuasion comparing to a situation that the robot is not present.
\end{enumerate}

This study used a repeated between-subject design with four conditions: Low Reward (LR), High Reward (HR), a control condition with 0 Reward (0R) and a condition with No interaction with the Robot (NR). It should be mentioned that in the last condition we used the low value of the reward.  
More specifically, we investigate the effect of repeated interactions within subjects. In addition, we investigate the effect that different level of exerted power may have between subjects. 

\subsection{Measures}
The participants were requested to fill out a pre-questionnaire including  demographics (Age, gender, Nationality, Occupation, and Field of study).
As we ran the experiment in English with mostly non-native English speakers, we asked participants to rate their English proficiency on a 5-point Likert scale (1 Basic - Professional 5).  
Previous studies indicated different attitudes among people who interacted with robots earlier. So, similar to previous studies, we checked if the participants had already interacted with robots in general, and if they had already interacted with Emys before this experiment. 

Next, the participants were asked to respond the Personal Sense of Power (PSP) questionnaire~\cite{anderson2012personal} that gives us an idea of their social power level. In addition, the participants were also requested to complete a short version of the Eysenck  personality questionnaire~\cite{francis1992development}, that gave us information on their levels of Neuroticism (N), Extroversion (E), Psychoticism (P), and Lie scale (L).  

Also, after finishing the task, we asked the participant to respond a post-questionnaire to have a better understanding of their perception. To measure how they perceived the robot, we applied the RoSAS questionnaire~\cite{carpinella2017robotic}. 
As the robot was giving rewards to the participants, we measured the extent to which this action gave the robot Reward Social Power. 
Finally, we asked them specifically if they changed the selected category at any iteration to make sure they understood the game. To better understand why they made such decisions, we asked them to clearly state why they have accepted/rejected the robot's offer (an open-ended question).  
Finally, as multiple numbers of factors contribute to the processing of persuasive messages, we use the Susceptibility to persuasion scale~\cite{modic2018we} to measure a relatively broad spectrum of factors leading to persuasion. 

To further investigate the interactions of the participants within this task, we added a number of questions to the pre- and post-questionnaire (Table~\ref{table:st3_TSQ}). To investigate if the interaction with the robot influences \textit{trust} and how they believe the robot would give them the reward if promised, we added a question in pre- and in post questionnaires measuring this. 
Next, participants were requested to indicate on a 5-point Likert scale how much they like Quiz-type games, and how often they go to Cinema. 

Furthermore, with items 6-10 in Table~\ref{table:st3_TSQ}, we specifically check how persuasive they found the robot, if the robot was trying to change their mind, if they were convinced to change or they felt compelled	to change their initial selections.	

\begin{table}
    \caption{Study 3 Task Specific Questions}
    \begin{tabularx}{\columnwidth}{llX}
        \hline
        \# & pre-/post Q &  Question\\
        \hline
    \cmidrule{1-3}
        1& pre- & In this specific game, if the robot promises you ``a reward" in the game, to what extent do you think the robot will give the reward to you?\\
        2& post-&Consider this specific game, when the robot promised you "a reward" in the game, to what extent did you think the robot will give the reward to you?\\
    \cmidrule{1-3}
        3&pre-&How much do you like trivia games and quizzes (In General)?\\
        4&pre-&How often do you go to the cinema?\\
    \cmidrule{1-3}
    6&post-&How persuasive did you think EMYS was?  (Not at all persuasive 1 - 5 Extremely persuasive) [Persuasion is an attempt to change somebody's opinion]\\
    7&post-&Emys was trying to change your mind.\\
    8&post-&Emys could convince you to change categories.\\
    9&post-&You felt compelled to change categories.\\
    10&post-&Changing categories was a good idea.\\
        \hline
    \end{tabularx}
    \label{table:st3_TSQ}
\end{table}

\subsection{Participants}
In this experiment, 118 people (54 females) participated voluntarily in the response of receiving cinema tickets. To recruit participants, we put several advertisements around the university, as well as the university's Facebook group.  
The participants' age ranges between 18 and 79 years old (28.6$\pm$16.9 and 1.6 S.E.). 
The participants signed an informed consent form before participating approved by the Ethical Committee of the University.

Then we randomly assigned the subjects to the four conditions of the study and counterbalanced the data to have approximately equal number of females in each condition [30 people in LR (13 females), 
30 people in HR (13 females), 
30 people in 0R (13 females), 
28 people in NR (15 females)].

\subsection{Procedure}
\subsubsection{Task, Robot, and Environment}
In the designed task, persuasion is operationalized within a game. The participants were asked to play a trivia game in three trials with different categories of questions. The game contains 6 categories ("Animals", "Arts", "Astronomy", "Geography", "Science", "Sport", "TV and Movie") and each category can be selected only once. Each category contains 5 questions and a correct answer to each question carries 1 point. The order of the questions in each category is the same for all participants to avoid the order effect on the responses. 

To provide the incentive of the games, cinema tickets are given to the participants depending on the scores they collect. The higher the score the more tickets they gain. Specifically, the participants could get more than one cinema tickets (up to three tickets) based on a pre-defined rule (the first 7 points in score grants a cinema ticket, each 8 more points lead to another ticket). In this task, we selected cinema ticket as the final reward, which is more valuable than the pen used in the second study. 
In each round of the game, the robot proposes two of the mentioned categories and the participant selects one (without seeing the contents). 

To have a better understanding of the user preferences, we ask them to define an ordering of the topics based on their interest or knowledge (after the re-questionnaire and before starting the game). Based on this preference, the highest rated option will be offered against the lowest. We expect the participant to select his/her own highest ranked and the robot tries to change his/her mind. The robot always offers an option which has not been selected by the participant. 

\begin{figure*}
  \centering
  \includegraphics[width=\textwidth]{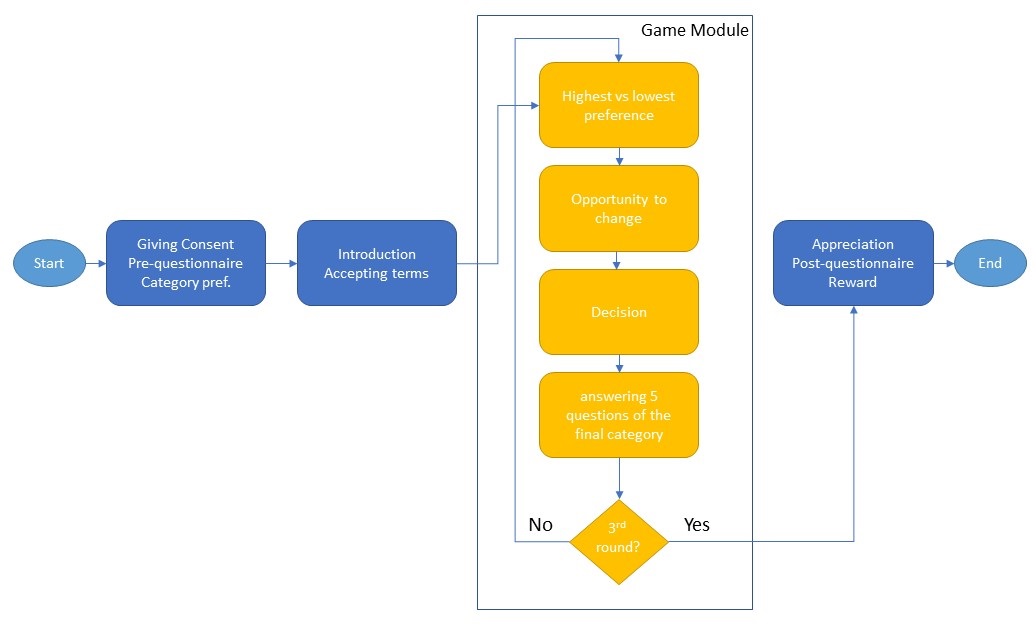}
  \caption{Study 3: Flowchart representing the study steps.}
  \label{fig:st3_gameflow}
\end{figure*}

For instance, if an arbitrary participant selects the following preferences: "Geography", "Science", "Astronomy",  "TV and Movie",  "Sport", "Animals", "Arts".
In the first round, s/he will be asked to choose from these two categories: "Geography" vs. "Arts".
We expect the participant to select the "Geography" category, as s/he indicated as her/his preference. And the robot asks the participant to change and select the "Arts" category.
This process is repeated for the other rounds (but the chosen category is removed from the list). The iterative manner design of the game gives us the opportunity to test H3. A flowchart of the full task is depicted in Fig.~\ref{fig:st3_gameflow}.

\begin{figure}
  \centering
  \includegraphics[width=0.75\columnwidth]{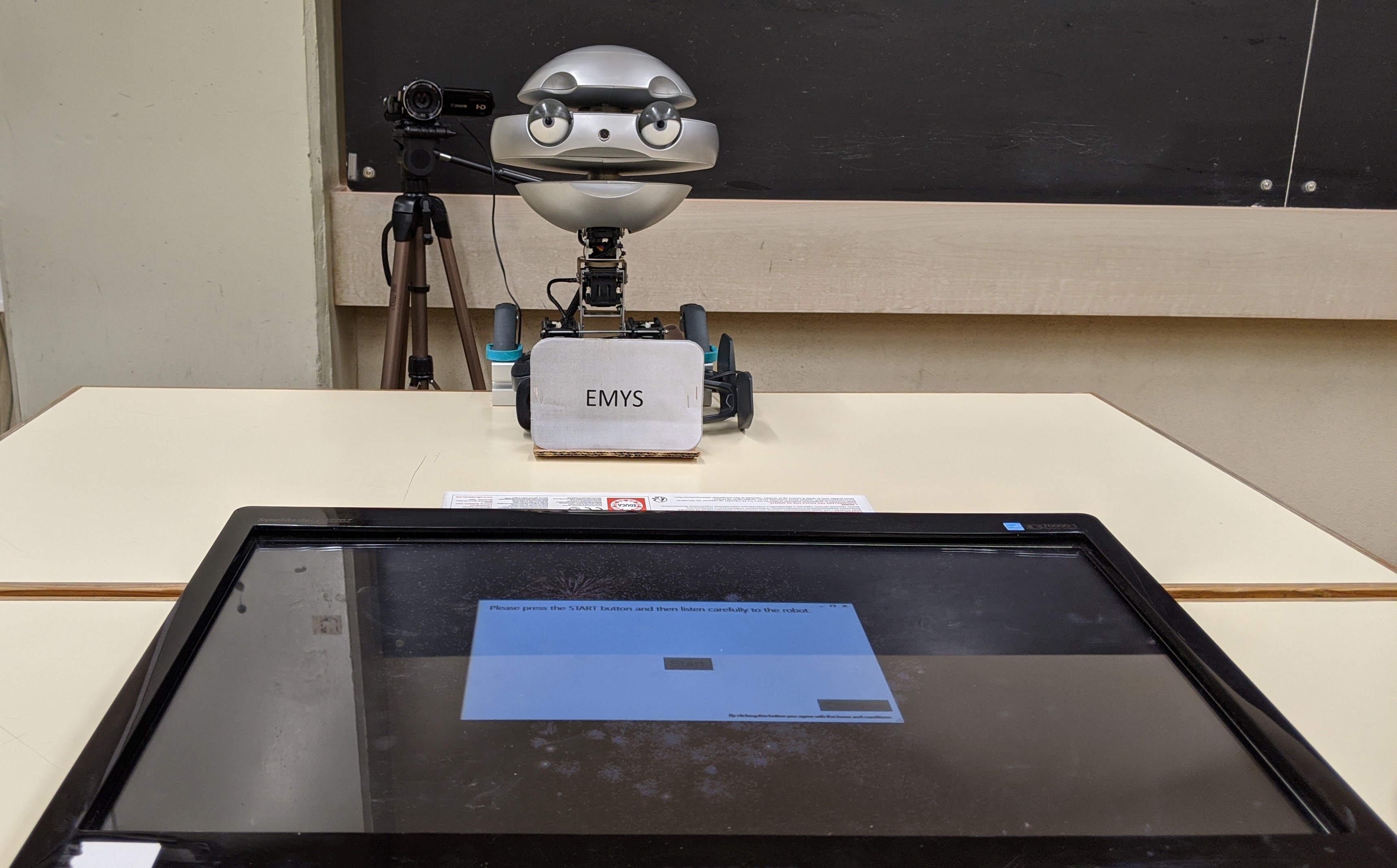}
  \caption{Study 3: Experiment Setup}
  \label{fig:st3_setup}
\end{figure}

In this task, similar to previous studies, we used the Emys robot mounted on a table in front of a touch-screen that is located between the subject and the robot (Figure~\ref{fig:st3_setup} depicts the study setup). 
The study took place in an isolated room. Each subject participated individually and during the game, the researcher stayed in the room to make sure no one cheats in the game (for instance, by searching the correct answers on the Internet).

The robot mediated the game by introducing the procedure and the scoring rules (introductory and ending dialogues are listed in Table~\ref{table:st3_dialogues}). 
Unlike the previous two studies, in this task the robot was fully autonomous (further details in Section~\ref{sec:st3_implementation}).

\begin{table}
    \caption{Study 3: main body of dialogues}
    \begin{tabularx}{\columnwidth}{llX}
        \hline
        Order   &  Category            & Dialogues    \\
        \hline
    \cmidrule{1-3}
        1& intro & [ANIMATE(joy4)]  Dear ``UserId''! Hello and welcome to this trivia game. \\
        2& intro &Let me explain you how the game works. I'm going to ask you a number of questions, categorized based on the subject.\\
        3& intro &You will select your preferred categories, in 3 trials.
        You can answer each category only once! Each category has 5 questions. Sounds good?\\
        4&  intro &You will quickly select what you think the answer is. Try to get as many right you can.\\
        5& intro &If your answer is correct, you will get X extra point(s). If your answer is wrong, you will not get any point. OK?\\
        6& intro  &At the end, if you succeeded to collect at least 7 points, [ANIMATE(joy1)] you're gonna win a fantastic prize! A cinema ticket! \\
        7& intro &And for each [emphasis level='strong'/] 8 more points, you will get another ticket! [ANIMATE(surprise4)] Exciting! right? \\
        8& intro &Now, let's start the game. Press [Gaze(button)] Continue if  you agree with the terms and conditions of the game. [Animate(joy1)]\\
    \cmidrule{1-3}
        19& final & OK. The game is over and you got + finalScore + points in total. Thank you very much for your participation and hope you have enjoyed the game! \\
        \hline
    \end{tabularx}
    \label{table:st3_dialogues}
\end{table}

\subsubsection{Implementation}\label{sec:st3_implementation}
\begin{figure}
  \centering
  \includegraphics[width=0.9\columnwidth]{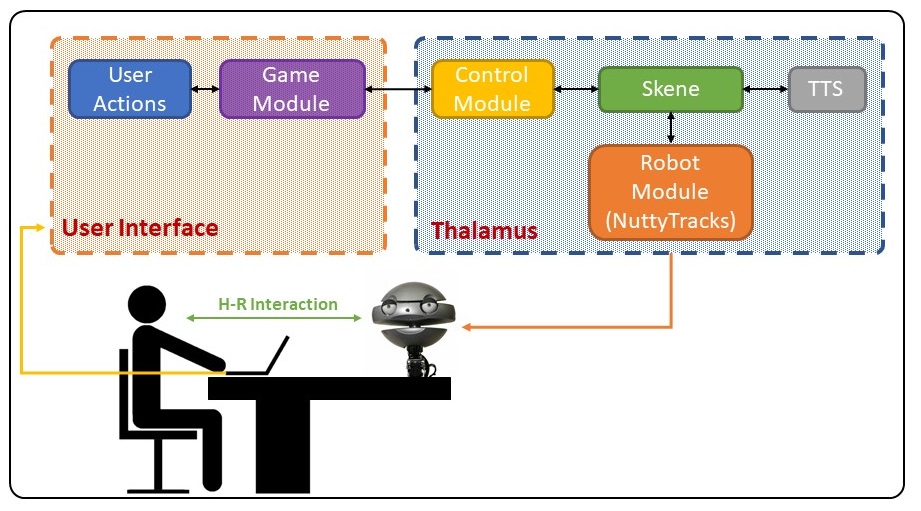}
  \caption{Study 3 - System Architecture}
  \label{fig:st3_arch}
\end{figure}

In this task, the robot performed in a fully autonomous manner. The core of our system architecture was the SERA Ecosystem~\cite{ribeiro2016ecosystem} which is composed by a model and tools for integrating an AI agent with a robotic embodiment in HRI scenarios. Figure~\ref{fig:st3_arch} shows the overall system architecture. We developed an application in C\# (displaying the game on the touch-screen and getting the answers of the participants), which is integrated with the decision state module using a high-level integration framework named Thalamus \cite{ribeiro2014thalamus}. This framework is responsible to accommodate social robots and provides the opportunity of including virtual components, such as multimedia applications \cite{ribeiro2014thalamus}. 

In addition, we used the Skene~\cite{ribeiro2014thalamus} behavior planner that provides the robot's behaviors, such as gazing, pointing, making speech, etc.
And a text-to-speech (TTS) component is used as a bridge to the operating system's built-in TTS. 
In the experiments using the Emys robot, we used a symbolic animation engine based on CGI methods called Nutty Tracks \cite{Ribeiro2013} which provides the capability to animate a robot in a graphical language.

A control module was developed to provide the communication between the display screen module, the Thalamus and the Skene. This module identifies and informs Skene, through Thalamus, about the utterance to be performed by the robot. Besides, this module controls the screen to be presented to the participant and processes the inputs made by him/her. With these inputs, the control module is able to send messages to Skene, that then sends to the robot module in order to perform the robot's animations. Hence, the overall interaction between the participant and the robot becomes fully autonomous in this way.

Since the persuasion happens in repeated interactions, we designed a pattern for the creation of dialogues to make sure the persuasion attempt is similar in all trials. In doing so, each persuasive message is consisted of four parts: 
1. Call for change,
2. Reward condition,
3. The goal,
and 4. Motivation. Table~\ref{table:st3_induction} lists the four parts of this pattern inline with the dialogues used in the experiment. Note that in this table, the first column represents the order of the dialogues in the final setting similar to Table~\ref{table:st3_dialogues} and~\ref{table:st3_random_dialogues}. The composition of these tables based on the order, composes the full dialogue used in this study. 

\begin{table}
    \caption{Study 3 induction dialogues}{In this table, ``C'' refers to the category selected initially by the user, ``O'' refers to the other category not selected, ``X'' refers to the extra point(s) that the robot offers as the reward, ``S'' refers to the current collected score, ``R'' refers to the required scores to a cinema ticket}
    \begin{tabularx}{\columnwidth}{llX}
        \hline
        Order   &  Category            & Dialogues    \\
        \hline
    \cmidrule{1-3}
        11& call for change&
           1. Now, I'm gonna offer you a chance to make a decision.\\
        && 2. How about selecting the other category?\\
        && 3. You chose the ``C'' category but, \\
    \cmidrule{1-3}
        12& reward condition & 
            1. If you prefer, I'll give you "X" extra points and you select the "O" category.\\
        &&  2. If you select the "O" category, I give you "X" extra point(s)! \\
        && 3. I'll give you "X" extra points if you select the "O" category.\\
    \cmidrule{1-3}
        13&the goal&
            1. Remember that you need at least 7 points to win a cinema ticket.\\
        &&  2. Up to now, you've collected ``S'' points. You need only "R" point(s) to receive a/another cinema ticket.\\
        &&  3. You are "R" points away to receiving a/another cinema ticket.\\
    \cmidrule{1-3}
        14&motivation&
            1. You will get closer to the cinema ticket with these extra points.\\
        &&  2. The extra points helps you to get closer toward a/another cinema ticket!\\
        && 3. Come on and select the other category to get closer to winning a/another ticket!\\
        \hline \hline
    \end{tabularx}
    \label{table:st3_induction}
\end{table}

The first part, ``call for change'', is the starting part of the persuasive message. First, the robot invites the participant to consider changing the selection, and next starts to influence him/her as follows. 
In the second part, the robot indicates the condition of giving the reward, i.e., if they change, they will receive some extra points (depending on the condition).
In the third part, the robot emphasizes the goal of the game and highlights the role of points in gaining cinema tickets. 
And finally, in the last part the robot motivates the participant to accept the reward and informs them the extra points could help them to win more easily. 
This pattern was checked by four researchers of this study and they agreed the creation pattern generates sentences with equivalent semantic loads. 
In this way, we perform a fixed level of induction in every persuasion attempt.

Apart from the persuasive strategy, the rest of repeated dialogues were displayed in a random order. This was done for two reasons: 1) to avoid repetition in scripting robots dialogues 2) also to have more diverse dialogues by combining different parts of smaller sentences. These dialogues are listed in Table~\ref{table:st3_random_dialogues}. 
For instance, when it comes to the decision making part of the game, the robot uses any of the three sentences of the first row. Exceptionally, on the very first trial the robot uses the first line, but on the second or third trial, it might use any of the remaining two on a random order avoiding repetition. 
Or, after each time that the user answers a question, the robot might use any of the ``gap fillers'' to start asking the next question.

\begin{table}
    \caption{Study 3 randomized dialogues}{In this table, ``C'' refers to the selected category, ``S'' refers to the collected score so far, \# refers to the number of question in the corresponding category.}
    \begin{tabularx}{\columnwidth}{llX}
        \hline
        Order   &  Category            & Dialogues    \\
        \hline
    \cmidrule{1-3}
        9&On decision making&
            1. [fixed in the first attempt] Alright! At this point, you have only two options[Gaze(options)]. You can select only one of these [Gaze(categories)] two categories.  Select your most preferred category on the touch screen.\\
        &&  2. Please select your preferred category once again!\\
        &&  3. Which category you would like to choose?\\
    \cmidrule{1-3}
        10 &After decision made&
        1. OK. Then let's start with category "C"! Get ready to answer! [Animate(joy1)] Here we go! [Gaze(button)] Click on the START button  when you are ready.\\
        && 2. Alright, [Animate(joy1)] then let's see the first question of category "C "! [Gaze(button)] Please click on START button to start with the first question!\\
        && 3. Alright[Animate(Animate(joy1))]! Ready for the first question? [Gaze(button)] Click on the START button!"\\
    \cmidrule{1-3}
        15 & Gap filler &    
             1. "Alright, next question.";
             2. "OK, next question!";
             3. "Next! ";
             4. "Question \#";
             5. "OK, next!";
             6. "Get ready for the next question!";\\
    \cmidrule{1-3}
            16 & On correct answers & 
            1. "Correct!";
            2. "That is correct!";
            3. "Your answer is correct!";
            4. "That answer is correct!";\\
    \cmidrule{1-3}
            17 & On wrong answers &
            1. "That is incorrect!";
            2.  "That is not true!";
            3. "Your answer is not correct!";
            4. "That answer is wrong!";\\
    \cmidrule{1-3}
            18&After each round &
            1. [1st attempt] "You have finished all the questions in this category. Your score is "S" up to now! Let's move to the next category!"\\
            &&2. [2nd attempt] Alright! Up to now, your score is "S"! Now, it is time to move to the last category!\\
        \hline
        \hline
    \end{tabularx}
    \label{table:st3_random_dialogues}
\end{table}



\subsection{Results}
Overall, 5 participants were excluded from the sample due to robot error. 
Before analyzing the data, we checked if there is any significant difference among the four conditions regarding any of the demographic variables. The results indicated that no significant difference exists between the samples in each condition regarding their age (F(3,119)=1.558,p=.203) and personality traits (N: F(3, 119) = .855, p=.467, E: F(3, 199)= .886, p= .451, P: F(3,119) = 1.011, p=.390, L: F(3,119) = 1.409, p=.244). 

Moreover, we verified that a prior interaction with \textit{robots} (t(85.595)=.972, p=.334) or \textit{Emys} (t(21.467)=1.011, p=.324) had no influence neither on the decision making of the participants nor their perception of the robot (Warmth: t(60.370)=-.559, p=.578; Competence: t(116)=.133,p=.894; Discomfort: t(166)=-.255,p=.799). 

Similar to previous studies, we investigated the results both objectively (participants' decisions to accept or reject the offer) and subjectively (task-specific questions). 
In the latter case, initially we checked the dimensionality of the scale using factor analysis for items 6-10 of Table~\ref{table:st3_TSQ}. The Cronbach's alpha indicated that removing item 7 increases the reliability of this measure (from .715 considering all 5 items to .779 when item 7 was removed). Hence, to measure persuasiveness subjectively, we averaged the 4 remaining items (6 and 8-10) that have more internal consistency.  
%




\subsubsection{Hypotheses Testing}
We investigated the first hypothesis both subjectively and objectively.
Considering the objective measure (decisions), having a binomial repeated measure among the independent groups (LR and HR), we analyze the data using Generalized Estimating Equations (GEE). With a significance of 0.901, there is not enough evidence to conclude that whether the higher reward has an effect on the outcome (being persuaded). 
%
Similarly, from the subjective perspective, the results of t-test indicated that no significant differences exist among LR and HR groups considering the persuasiveness score of the robot (t(55.913)=-.567, p=.573). 
Hence, we cannot verify the first hypothesis (H1).

To investigate the second hypothesis, we added having/not having interactions with robots as another predicting factor of the GEE model. However, with a significance of 0.825, there is not enough evidence to conclude that whether having an earlier interaction with robots has an effect on the persuasion. Furthermore, adding this factor increased the QIC (Quasi Likelihood under Independence Model Criterion) value, which also endorses that this item is not a good predictor for the model. 
To check this hypothesis subjectively, we added having/not having interactions with robots as a covariate to ANCOVA analysis. The results indicated that the covariate is not a significant predictor of perceived persuasiveness (F(1,57)=.438, p=.511, $\eta^2$=.008).
Hence, having prior interactions with robots does not affect the decision making of the participants and we reject H2.

\begin{figure}
  \centering
  \includegraphics[width=\columnwidth]{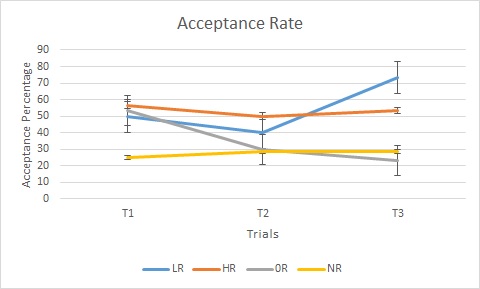}
  \caption{Study 3 - Percentage of acceptance in each trial}
\label{fig:st3_rate}
\end{figure}

With the third hypothesis (H3), we postulated that over repeated interactions, the effect of social power on persuasiveness does not change (decays/grows), considering that the level of power is fixed. We note that as the subjective measure was applied only at the end of the test, we cannot check this hypothesis subjectively.
And, we can only investigate this hypothesis objectively. 
In doing so, we included the trials as a factor in the GEE model. The results indicated that the repeated interaction has an effect on the first and third conditions, i.e., LR and 0R. To be more specific, on average, people in LR group were more likely to accept the offer at the third trial compared to the first trial (Wald(1)=4.807, p=.028). On the contrary, people in the 0R condition were less willing to accept the offer at the third trial compared to the first trial (Wald(1)=5.703,p=.017). 
Figure~\ref{fig:st3_rate} represents these findings. The acceptance rate decays only in the control condition, in which no persuasion is exerted. The rate stays unchanged in HR and NR conditions, but gets an increase in the LR condition. Hence, the effect does not decay and under specific conditions it grows over time. Hence, we accept the third hypothesis.
We would also like to highlight that the NR has lower acceptance rates, and the persuasion of 0R decreases, while with LR or HR it does not decay.


The fourth hypothesis (H4) investigates the robot's perception based on the RoSAS questionnaire. With this hypothesis, we expect to observe a higher score of \textit{warmth} in conditions that the robot gives rewards to the participants (persuasion conditions, LR and HR). The results of ANOVA test indicated that, although  in HR condition, the robot was scored higher on warmth and competence compared to the other groups, however these differences were not significant (Warmth: F(3,114)=2.174, p=.095; Competence: F(3,114)=2.299, p=.081; Discomfort: F(3,114)=.395, p=.757).
Hence, we reject the fourth hypothesis.

Similar to previous hypotheses, we checked the fifth hypothesis both objectively and subjectively considering LR and NR groups that differ only in one factor, which is the function of the robot. From the objective point of view, GEE indicated a significant effect of robot presence on the decision making of the people in LR group in comparison to NR (Wald(1)= 7.838, p=.005). 
Furthermore, people in LR group are more likely to accept the offer of the robot (Wald(1)=10.759,p=.001).

From the subjective perspective, the result of a t-test indicated a significant difference between the score of persuasiveness of the robot (t(56)=2.461, p=.017) and the higher mean in LR condition indicated that people found the robot more persuasive than the computer application (M=3.3167, S.E.=.19288 in LR vs. M=2.6339, S.E.= .19940 in NR).
Hence, the results verify H5, i.e., although the robot did not have any physical interaction, but its presence itself leads to higher persuasion.

\subsubsection{Exploratory Findings}
Apart from the postulated hypotheses, we further investigated the data to have a better understanding of the interaction under these conditions. 
As mentioned earlier, apart from the standard questionnaires, we added 5 other questions to the post questionnaire, specifically designed for this task (Table~\ref{table:st3_TSQ} item 1-5). 
For instance, the first two items measure the trust before and after the interaction. As the interaction with the robot and receiving the rewards (in reward conditions) might influence the trust indirectly, we checked if participants' trust in the robot (or how much they believe the robot fulfills the promised reward) remains unchanged during the interaction. The result of a paired sample t-test indicated that there is a significant difference between the scores of trust before and after the interaction (t(117)=-2.854, p=.005). And the higher mean after interaction indicated that the trust increased after playing the game (trust score before interaction: M=3.64, S.E.=.089 vs. after the interaction: M=3.86, S.E.=.092).
A post-hoc analysis indicated that this difference is only significant among the participants of HR group (t(29)=-2.362,p=.025) [M:3.46, S.E.=.18; M:3.83, S.E.=.19]. 

To control the bias of the trust change, we need to include it as a covariate. However, in order to add it as a covariate in GEE (regarding decision making), it should have been measured at each trial, while we had measured it only on the first and the last trials. To handle this, we averaged the trust score before and after and considered it as the trust score of the middle trial. And then we included these three scores as a covariate in the GEE model. This covariate increased the goodness of fitness, meaning that the new model is less fitted to the data. In addition, it was not a good predictor of decisions (Wald(1)=.523, p=.469).
Hence, there is not enough evidence if the trust factor was a good predictor of the model. However, this might have happened due to the missing measurement of trust in the middle trial. In other words, the average score might not be a good estimation of trust in the middle score. To overcome this doubt, 
we put aside the HR condition (as it was the only groups  with significant differences in trust scores before and after the interaction) and skipped the trust factor which was not statistically different  among other groups. The results indicated that the goodness of fitness of the new model was lower than the previous one, that is to say, we achieve a better model without HR. 

To check this intervention subjectively, we added the trust difference between before and after interaction as a covariate in ANCOVA to check its potential influence on perceived persuasiveness. The results indicated that there is no statistically significant difference between adjusted means of persuasiveness with regard to trust difference (p=.654, effect size .004).
In sum, although the trust in the robot was increased after the interaction, this factor did not significantly influence the persuasiveness.

We further checked if personal preferences (i.e., liking trivia games or cinema) have affected the results. From the objective perspective, the results indicated no significant effect of liking cinema (Wald(4)=4.176, p=.383). However, liking trivia games turned out to be a good predictor of behavior (Wald(4)=9.671,p=.046). In other words, the more the participants liked trivia games lead to a higher likelihood of accepting offers (Wald(1)=5.594,p=.018). 
From the subjective perspective, we performed ANCOVA with \textit{liking quiz} as a covariate. The results indicated that this covariate does not significantly predict the dependent variable, i.e., persuasiveness of the robot (p=.827, effect size: .000)   

Furthermore, earlier research indicated that when there is a power match between the persuader and the persuadee, higher persuasion is achieved~\cite{mourali2013powerful}. To investigate this effect, we labeled the participants as high/low power based on their PSP (Personal Sense of power) scores (the ones scored higher than the average were labeled as high power). Moreover, we labeled the robot as being high/low power based on the scores associated with the reward social power (higher than the average score was labeled as high). Then we checked if a power match exists between the participant and the robot and included it as a covariate in an ANCOVA analysis~\footnote{We would like to highlight that scores of power match were equal in case of median- and mean-split.}.  
The results indicated that the covariate does not adjust the association between the predictor and the outcome variable. 


Finally, in~\cite{ghazali2018influence} Ghazali et al. considered the total number of accepted offers as an indicator of compliance.
Similarly, to check if this feature is a predictor of behavior, we applied an ANOVA test and the results indicated a significant difference among the four conditions (F(3, 114)= 4.682, p=.004, $\eta^2=.110$).
Post-hoc comparisons using the Tukey HSD test indicated that the mean score for the LR condition (M=1.633, S.E.=.183) was significantly higher than NR condition (M=.821, S.E.=.189).
Similarly, the mean score in HR condition (M=1.600, S.E.=.183) was significantly higher than NR.
This results endorse the verification of H5, in other words, the presence of the robot has significantly affected the decision making of the participants.
%

From the subjective point of view, the result of a Pearson correlation test indicated a strong, significant, and positive correlation between the total number of accepted offers and perceived persuasiveness. Particularly, the higher the perceived persuasion, the higher the number of accepted offers (r(118)=.679, p=.000). 
Further analysis indicated that this correlation is stronger among LR (r(30)=.795, p=.000), then 0R (r(30)=.781,p=.000), then HR (r(30)=.490, p=.006) and the least for NR (r(28)=.443,.018).
This finding is inline with~\cite{ghazali2018influence}, that when the persuasion is stronger, the compliance decreases due to potential reactance. 






\subsubsection{Analyzing the Game Log}
Apart from the data obtained from the questionnaire, we analyzed the game logs to further investigate how the participants acted and made decisions during the game. In this subsection, we investigate a number of these features. 


One of the game features that might influence the decision making of participants to accept/reject the offer is the remaining scores they require to achieve a cinema ticket. Adding this feature to the GEE model indicated that  
it is a good predictor of the behavior (Wald(1)=6.386, p=.012) and the test of goodness of fit showed a decrease (470.024 vs. 476.080) in the Quasi Likelihood under Independence Model Criterion (QIC) meaning that it is a good predictor of decisions. In addition, the results indicate that an increase in distance to a ticket increases the likelihood of compliance (log odds: B: .116$\pm$.0458, Exp(B)=1.123).

Another fruitful feature of game logs might be the collected score in each trial. The test of model effects indicated that this is also a good predictor of decision making (Wald(1)=27.409, p=.000). Additionally, the goodness of fit test showed a decrease in  QIC (451.646 vs. 476.080), meaning that a model using this factor as a predictor is a better fit to the data. On the other hand, further analysis indicated that 
an increase in the score leads to a lower likelihood of acceptance, i.e., the lower the score they collected, the higher the probability of accepting the offer in the next trial (odds: -.444$\pm$.0848; EXP .642).

Based on this, we grouped the observations by the current score at each round (as an indirect measure of distance to the ticket) and compared the persuasiveness score of the robot considering the condition as the covariate. The results indicated no significant difference among the scores of persuasiveness with regard to the score (F(8,119)=.369, p=.935).
However, regarding the objective measure (decisions), there was a significant difference between the decision making of the participants only on the first trial (T1: F(8,119) = 2.224, p=.031). The result did not yield to any significant difference on the second (T2: F(8,119) = 1.241, p=.282) or third trial (T3: F(8,119) = 1.826, p=.079).

Another game feature that might be informative is the cumulative score or the overall score they gained before making a decision. The test of model effects indicated it as a good predictor (Wald(1)=9.760, p=.002) and the  goodness of fitness test showed a decrease in QIC (460.854 vs. 476.080) further endorsing it is a good predictor. 
However, the negative log odds, i.e. -.226$\pm$.0724 
indicated that people having collected high scores are less likely to accept the suggestions. 

In this game, the participants could change their mind and select options contrasting their stated preferences during the game, if they understand the strategy of the robot. This feature gives them an opportunity to cheat the robot. In other words, as the robot always offered the opposing option (the option that participants did not choose), they could initially select the option that they do not prefer, and accept the suggestion of the robot to answer their most preferred option (plus receiving points in persuasion scenarios).
In this regard, it is not unlikely to observe less cheating on the first trial, when the participants are not aware of robots' function. However, the cheating might reach to its maximum on the third trial when the participants become more familiar with the robot's function. 
Analyzing the game log indicated that only one person selected his nonpreferred choice and changed to the preferred option in all trials. In addition, two people did this both in the second and the third trials, however, one of them belonged to the 0R condition (s/he might do this randomly), and one belonged to NR. 
Hence, although the participants had a chance to cheat the robot, only two persons did this (one in HR and one in NR).


\begin{table}[!t] 
\caption{GEE model summary for decision change direction}
\label{table:st3_GEE_dirchang} 
\centering 
\begin{tabular}{ccc}
\toprule
        \hline
        Conditions   &  Odds            & Wald test result    \\
        \hline
    \cmidrule{1-3}
    \multicolumn{3}{c}{trial 1}\\
    LR vs. NR & odds:0.758&      Wald(1)=3.983,p=0.046\\
    HR vs. NR & odds:0.875&      Wald(1)=5.655,p=0.017\\
    0R vs. NR & odds:0.758&      Wald: 4.212,p=0.040\\
    \cmidrule{1-3}
    \multicolumn{3}{c}{trial 2}\\
    \multicolumn{3}{c}{No significant differences found}\\
    \cmidrule{1-3}
    \multicolumn{3}{c}{trial 3}\\
    LR vs. 0R & 1.253 &       Wald(1)=8.860,p=0.003\\
    HR vs. 0R & 1.022 &       Wald(1)=4.996,p=0.025\\
    NR vs. 0R & 1.232 &       Wald(1)=4.326,p=0.038\\
\bottomrule
\end{tabular}
\end{table}

Finally, we investigated how the participants made decisions considering their initial preferences. In this regard, they could change their opinion in three possible directions: not accepting the offer or no change, accepting the offer and select a less-desirable choice, accepting the offer and select a more desirable choice.
Having a repeated test, we applied GEE model using change directions as responses and trial and conditions as predictors. Table~\ref{table:st3_GEE_dirchang} summarizes the findings.

\begin{figure}
  \centering
  \includegraphics[width=0.95\columnwidth]{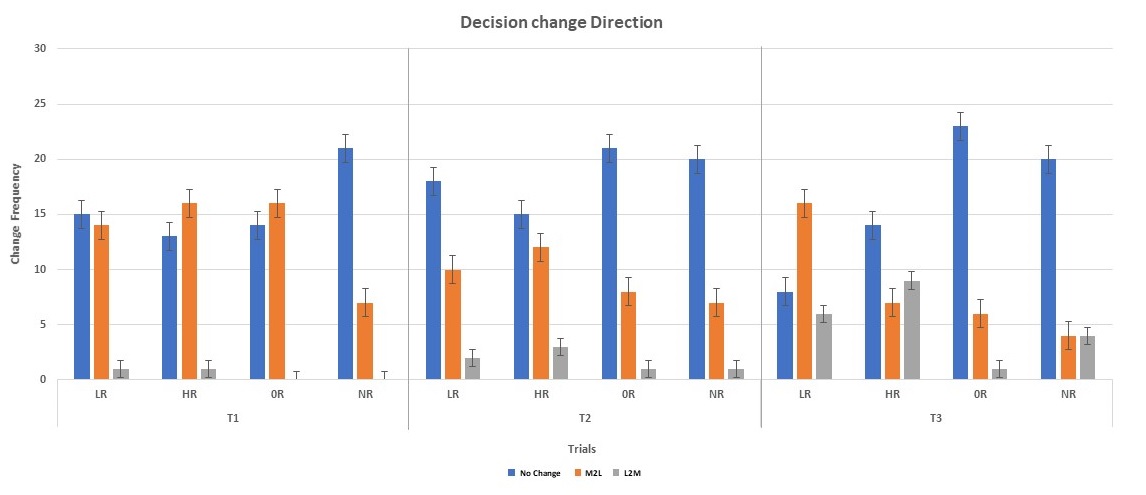}
  \caption{Study 3 - Direction of decision change} {In this figure, blue bars represent no change (rejecting the offer), M2L stands for changing from a More favorable choice to a Less favorable one, and L2M represents the contrary. }
  \label{fig:st3_dirchange}
\end{figure}

In sum, we can infer that, in the first trial, the robot was effective and all conditions having the robot were higher persuading, even the one with 0 reward. In other words, on the first trial, the robot could persuade the users to change their initial selection significantly more frequently than the NR condition. 
In the second trial, the differences between groups were not significant. There is not enough evidence to make any conclusion on this trial.
However, on the third trial, all conditions are higher than 0R, meaning that the reward has an effect on decision making (manipulation checked). Here is the only place were NR become more persuading than 0R. And when the robot is not giving any reward and does not have any power on the participants, over a repeated interaction, it acts less persuading over time in comparison to a computer application giving reward constantly (Figure~\ref{fig:st3_dirchange}).

A potential reason might be the differences in their selection during the game and what they stated in the initial questionnaire. As we mentioned earlier, it is not probable that the participants wanted to cheat the robot. They might have not paid enough attention while they were filling out the preference questionnaire, but when it came to reality in the game, they paid more attention to what they prefer more. 

In the LR condition, people were struggling to get at least one ticket, while in the HR condition, people mostly received one ticket and were struggling to achieve an additional ticket. Maybe in LR one extra point could help them to win at least one cinema ticket, but in HR there was a different situation.


 
\subsubsection{Qualitative Analysis}
To further investigate and learn more from the data, we took a closer look by revisiting the data qualitatively. With this new view, we aim to interpret and describe data to find and understand potential patterns of behavior in the decision making of the participants.
That is to say, the open-ended question aimed at exploring participants' unique perspectives and motivation behind their decision making.

As the result of a qualitative analysis is directly dependent on the coding scheme used, three researchers contributed in coding (data retention) to have a more credible and reliable coding scheme~\cite{richards2014handling}.
Initially, the three researchers labeled the data freely and individually to achieve categories as general as possible and not to get biased by each other. We used the WebQDA software~\footnote{https://www.webqda.net/} to code and label the open-ended question. Then they discussed and compared the labels to enhance the validity of the coding.

Each of the three researchers identified 13, 11, and 9 categories (listed in Table~\ref{table:st3_p_codes}) in the first phase.
To determine the trustworthiness of the codes, we further discussed them through checking the consistency and reasons for inconsistency. We reached an agreement of 8 categories in the end (low self trust, 
unknown,
reward,
high self-trust,
Emys,
non-compliance,
good offer,
game experience).
In the next step, we relabeled the data using this coding scheme with the goal of minimizing inconsistency. The inter-rater reliability between each two researchers gained K=.653 (p=.000), K=.515, p=.000, and K=.476, p=.000.  
After another discussion, we agreed to use one of the ratings with the highest inter-rater reliability. 

\begin{table}
\centering
    \caption{Study 3 - Primary coding scheme by 3 researchers}
    \begin{tabular}{lll}
        \hline
        Researcher 1   &  Researcher 2            & Researcher 3    \\
        \hline
    \cmidrule{1-3}
        Challenge	&Reward	&Bad Offer\\
        Curiosity	&Self trust 	&Challenge\\
        Emys	&Low self trust	&Confused\\ 
        Fun	&Challenge	&Curiosity\\
        Lack of knowledge 	&Game experience	&Good Offer\\
        Low reward	&Good offer	&More Credit\\
        Misunderstood	&Non compliance	&Persuaded\\
        No reason	&Emys	&Random Choice\\
        Distrust	&Unknown	&Self-Trust\\
        Points	&Other&	\\
        Risk	&Low reward&\\	
        Self-trust		&&\\
        Unknown		&&\\
    \cmidrule{1-3}
    \hline
    \end{tabular}
    \label{table:st3_p_codes}
\end{table}

%


\begin{figure}
  \centering
  \includegraphics[width=\columnwidth]{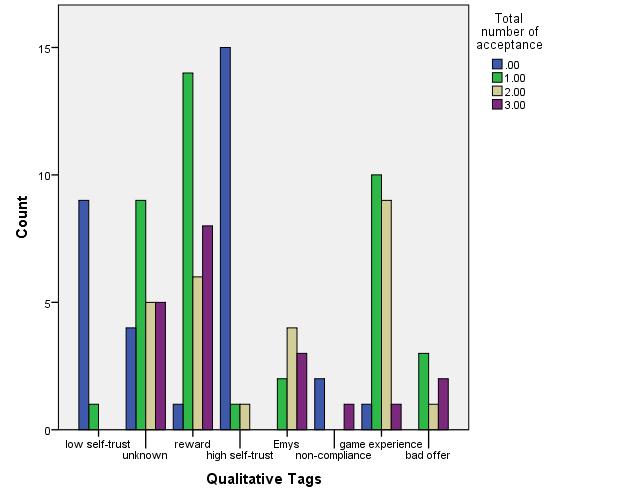}
  \caption{Study 3 - Distribution of total number of accepted offers over Qualitative Tags}~\label{fig:st3_q_acc_tot}
\end{figure}

\begin{figure}
  \centering
  \includegraphics[width=\columnwidth]{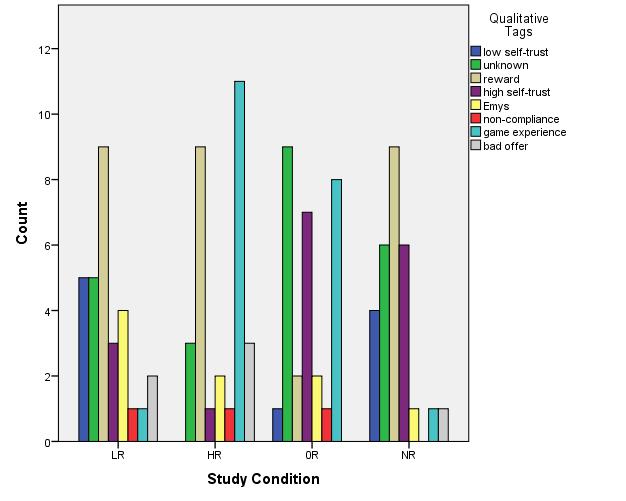}
  \caption{Study 3 - Distribution of Qualitative Tags over conditions}
  \label{fig:st3_qualitative_tags}
\end{figure}

We analyzed the collected data using the coding scheme and the findings are summarized as follows:
\begin{enumerate}
    \item There is a statistically significant difference in the number of accepted offers between people categorized with regard to different tags (f(7,117)=10.160, p=.000). People who had high self-trust were less compliant. These people who valued their own knowledge than the robot's offer tended to reject more than the rest (Figure~\ref{fig:st3_q_acc_tot}).  

    \item Participants' perception of the robot was significantly affected by their statement regarding robot persuasiveness (F(7,117)=7.924, p=.000) but no difference exists in RoSAS scores (Warmth: F(7,117)=.787, p=.599, Competence: F(7,117)=.678, p=.690, Discomfort: F(7,117)=.559, p=.787).

    \item The distribution of the tags is significantly different in different conditions ($X^2$(21)=41.248, p=.005), as depicted in the Figure~\ref{fig:st3_qualitative_tags}. A Bonferroni post-hoc tests indicated that these difference stands out regarding the game experience tag among HR vs. NR and HR vs. 0R. Specifically, people in HR condition indicated that they accepted the offer to have a higher game experience by accepting the challenge to answer their less-desirable choice. And this higher number (11 in HR and 1 in NR and 1 in 0R) is statistically significant. 



\end{enumerate}


Apart from this quantified reasoning, Figure~\ref{fig:st3_qualitative_tags} depicts that LR and NR (conditions with the same reward level, i.e., 1 extra point) have the same number of low self-trust, and all conditions have the same number of reward, except 0R. Furthermore, as discussed earlier, the quantitative results could not verify H1 (i.e., no significant difference between HR and LR regarding decision making). 
Interestingly, we observe a higher number of people with low self-trust in the first condition (LR).
The responses to the open-ended question indicate that people in the high reward condition (HR) were less compliant and that could be related to the fact they wanted a challenge, and maybe they were more risk-prone because they did not have to worry too much about the ticket (they had more freedom to change or not and still secure the ticket). 

In addition, a closer look to the data indicates that in the first control condition (0R) on the first trial, people tend to accept, because they expected to receive some points. For instance, an individual indicated that ``I hoped  it asks easier question when I changed the subject.'' But when they received none, they stopped accepting the offer (for example, as indicated by the individual ``When I did, I did not change anymore''). Therefore, after not receiving any points, they stopped accepting the offer.

Overall, the qualitative analysis opened up new insights to the data. As ``no qualitative methodology is exclusive''~\cite{richards2014handling}, we do not claim that the coding we used is the only applicable scheme.  
Although three researchers coded the sentences individually and in a group to enhance content validation; however, as one's coding changes over time~\cite{richards2014handling} there might be other interpretations of qualitative results.
Hence, these findings should be considered cautiously. Furthermore, apart from the coding process, some  participants might be shy to directly indicate they wanted the ticket, i.e., the reward.




\subsection{Discussion}



In this section, we presented the results of a user-study performed to investigate the effect of different levels of social power (particularly reward social power) and repeated interactions on persuasion. 
We hypothesized that a higher level of social power would lead to higher persuasion, and having a fixed level of social power, this effect would not decay over time. 
The result of this study did not verify the former neither subjectively nor objectively, and hence, we could not conclude if the increase in power leads to higher persuasion. This finding is  similar to the results of the second study, in which the increase in ratings (that indirectly increased the level of reward) did not lead to any significant difference in the decision making of participants. Hence, we may conclude that persuasion does not have a linear relationship with the level of power exerted. This is inline with recent research that indicated a nonlinear relationship between power and persuasion~\cite{dubois2016dynamics}.

On the other hand, Ghazali et al. endorsed that exerting a strong persuasion attempt acts negatively and hence causes reactions and leads to low compliance~\cite{ghazali2018effects}. 
Also, they indicated the reaction is associated with higher negative cognition and feeling of anger, which might be equivalent to a higher score of discomfort dimension of the RoSAS questionnaire. However, our results did not lead to any significant differences in the score of discomfort for people who rejected more frequently compared to others (ANOVA: F(3,114)=1.330, p=.268).
In this case, although in HR condition the persuasion was stronger, reactance has not happened. In other words, the rejection was not due to any reaction felt measured by the RoSAS questionnaire. 

Hence, our study verifies that power and persuasion do not have a linear relationship, however, further investigation is  required to determine this nonlinear relationship. In addition, further evidence is required to assess the reactance threshold. 
Apart from this, another potential reason for this finding might be the small difference between the scores in LR and HR conditions. Although we considered the higher reward to be more than half of the maximum potential achievable score (3 out of 5), participants might have valued this extra score different from our expectations. A clear information about the state of their mind might be a clue to interpret the results.


Further, the results lead to a contradicting finding regarding the latter, i.e., repeated interactions. 
Specifically, although we expected that the effect of power on persuasion remains unchanged over a repeated interaction, this hypothesis was verified only in two conditions, HR (high reward) and NR (no robot).
Particularly in LR (low reward) and 0R (zero reward), the persuasion was not the same on the three trials. 

In the case of 0R or the first control condition, people tend to accept the suggestion to change less frequently at the third trial. When they were not gaining any scores for changing, they trusted their own knowledge and did not accept. Hence, not using any sort of power strategy, the robot did not have any persuasive power and people did not comply with the request.
However, despite our expectation, in LR condition, using the same level of power, the robot gained higher persuasion at the end. And in this case, the persuasion was even higher than HR condition.
Interestingly, this finding is inline with Ghazali et al.~\cite{ghazali2018effects} that the robot with mid-level of persuasion power was more successful than high-power or not robot. Although earlier, the result indicated no reaction in HR.

We argue that this inconsistency may be due to the value of the reward that the participants associated with in each trial. In other words, it seems possible that the value of the reward was not equal in all conditions. That is to say, when the participants were more near to gaining a cinema ticket, a single score might have a value more than one score in the first trial when they are far away from getting a ticket. As an example, imagine a participant needing only one score to gain a cinema ticket. This one single score means more to him/her, compared to a person needing 5 scores. 
However, as reported earlier, based on the collected data, we did not find any significant difference of persuasion between groups of people considering their scores. To

Furthermore, in spite of what we hypothesized, people new to robots did not show a significant different behavior compared to the others. This finding is contradicting to Study 2 findings.
A possible explanation for this might be the small sample size of study 2. Around one-fourth of the sample of study two were new to robots and they mostly fell in the same condition. Similar to the previous study, 33\% of sample had already interacted with the robots. However, in this study, not only the sample size is doubled, but also the sample was more uniformly spread in the groups (each group had around 70 percent people new to robots, except in the RL group that 57 percent of the sample were new to robots).

Unlike the two previous experiments, we did not find any significant differences in the perception of the robots.
In the first study, the two robots used two sets of completely different dialogues in their interactions with the participants. Also, in the second study, the robot used two different strategies and correspondingly different dialogues in the persuasive strategy. 
However, in this study, the difference between the conditions was minor and only one single strategy was used in the persuasion conditions. In addition, the reward did not increase the likeability of the robot. 
Hence, we rejected the fourth hypothesis.

Finally, we considered the fifth hypothesis (H5) to investigate if the presence of the robot has any effect on the persuasion. Specifically, one might argue that since the robot has no physical interaction with the participants, the persuasion is gained only due to the scores that people receive. In other words, a sole application would do the same job. 
This is the main reason for adding the fourth condition (NR).
However, the results indicated that this argument is not true and the robot attendance leads to another channel of persuasion due to its social presence. Hence, the sense of presence of the robot should not be neglected in this case. This finding is inline with Ghazali et al. ~\cite{ghazali2018effects} as well. Although their results indicated that in one condition, i.e., low psychological involvement, the increase in social agency did not influence compliance, but in another condition, or high psychological involvement, compliance remained the same for the medium social agency but dropped for the high social agency condition. 
However, our finding shows a different trend, i.e., when there was no robot, we have achieved significantly less compliance. A potential difference might be that in our study the robot was present, but turned off. Although most of the participants thought ``Emys'' was the application, we made the expectation that they will interact with the robot later on in another phase shortly after finishing the current task. 

Importantly, we would like to highlight that the results indicated no significant difference between the two control groups, i.e., 0R and NR, in which there was no manipulation~\ref{fig:st3_rate}. This finding further highlights the effect of the manipulation we made. Specifically, this finding presents consistency and decreases the probability of unobserved bias in the data due to selection.


%




\subsection{Summary of Findings}
The findings of this study are four-fold: first, inline with other studies, the results indicated that an increase in reward social power (and more generally a stronger persuasion) does not necessarily lead to a higher compliance. In other words, a robot with medium level of power could be more persuasive than another with higher power.

Second, a prior interaction with robots does not influence the decision making of the participants, unlike what we observed in the second study, which might have occurred due to the smaller sample size and using a single persuasion attempt. 

Thirdly, over repeated interactions, the compliance might change due to the specific circumstances (either the study condition or the user's valuation of the reward) of the study. However, further evidence is required to determine how these circumstances affect decision makings. 

And finally, the qualitative analysis of the contextual data gathered in the study revealed new insights to the data. For instance, people with high self-trust were less compliant with the robot. In addition, people in the HR condition felt a higher game feel and were more willing to challenge by accepting the offer.  Furthermore, in the 0R condition, people expected to gain something by accepting the offer in the first trial. And that is why they were compliant with the robot.



\subsection{Limitations and suggestions for future studies}
One limitation of this study is the use of the questionnaire only before and after the study. In other words, we do not have enough information about the user at each single trial. Hence, we could not measure the subjective factor (robot perception regarding persuasiveness or RoSAS). Furthermore, we did not have enough information about how they perceived the trustworthiness of the robot on the second trial.

Another potential limitation of this study is  the unknown value of each single point to the subject. As discussed earlier, 1 single score might have a different meaning to each individual. 
This becomes particularly important considering the cumulative scores over the three trials mentioned earlier. Thus, maybe at each trial we can reset the scores, so that the next round would be independent from the distance to score. Or, we may have a large pool and compare people with the same amount of remaining scores separately. Or more ideally, we may inquire more subjects to check how valuable they find each single point. 

Like any other self-report measure, the primary questionnaire asking about the preference might not be a good measure of users' preferences. In fact, some people selected their less favorable choice initially and indicated in the open-ended question that they did not indicate their preference carefully before the game. Hence, considering that there may be a cheating incentive, we cannot make sure if they really selected their preferences carelessly or they decided to cheat.

Although discomfort is supposed as an indicator of negative cognition, but it might not be a good predictor of reactance. For instance, different evidences indicated less compliance in HR than LR. This might have happened due to reactance to the robot's suggestion when using a higher level of power or that might be due to the remaining score to get a cinema ticket. As we discussed earlier, the analysis of the game log indicated that at the third trial, people in LR accepted the robot's suggestion more frequently than HR. 
The study is limited by the lack of information on reactance and a better measure is required. 
In addition, if the evidence suggests the occurrence of reactions, a future study could assess the effect of different power levels to indicate the level threshold at which reactance happens. In other words, considerably more work will need to be done to determine the relationship between power level and persuasion with regard to reactance. 

Further research could also be conducted to determine the effectiveness of behavioral analysis of the participants using the recorded videos. Apart from the contextual data that we analyzed earlier, these behavioral cues could enrich the qualitative analysis. 

In final words, these findings provide insights for future research that reward social power endows persuasiveness to robots. Further work needs to be done to establish whether other power bases are effective in persuasion.

\section{Conclusion}


In sum, our contributions provide new empirical findings and design implications for using social robots in the compliance gaining and behavior change context. 
Specifically, the link between power and persuasion investigated in this work may contribute in addressing some HCI/HRI research problems~\cite{hashemian2021thesis}.

Using social power bases, we attempt to design social robots, a specific case of social agents, equipped with social power bases. We selected social robots due to their physicality and the higher sense of presence compared to virtual agents.

We operationalized these power bases within persuasion tasks and attempt to investigate this potential application of social power in human-agent interaction, i.e., persuasion. The link between power and persuasion, as well as the recent application of persuasive technology, motivated us to investigate this link further. 
We designed three persuasive strategies inspired by social power, particularly expert, reward, and coercion. 

Together, the results of these studies provide important insights into persuasion in HRI.
We argue that our contribution advances the study of robot persuasion by testing new factors (social power strategies) that may affect persuasion effectiveness. In this direction, our contributions are as follows:

\begin{itemize}
    \item We identified that the use of social power by social robots is effective for persuading people.
    \item We investigated this effect using incentivized real choice and nonimaginary tasks that increase the external validity of the design.
    \item We used different within- and between-subject studies as well as mixed-designs and investigated the power-persuasion link both objectively and subjectively in the three studies.
    \item We concluded that one strategy could influence the users both objectively and subjectively. And these two channels of persuasion might not happen both at the same time (as observed in the first study).
    \item We argue that social rewards can be effective at persuading users and, unlike material rewards, they are unlimited and always available at a lower cost.
    \item We observed that people who are new to robots might be affected by the novelty effect and this threatens the external validity of results. In this case, a longer interaction might mitigate this effect (Study 2). 
    \item We found that to achieve a significantly different perception of the robot in case of warmth, competence, and discomfort, the robot dialogue and social cues should be notably different. In other words, minor differences in dialogue sentences might not lead to a high difference in these scores (as observed in the third study). 
    \item Having a fixed level of social power, the effect of power on persuasion does not decay over repeated interactions (Study 3). The effect might become stronger under specific circumstances.
    \item An increase in the level of power does not linearly give rise to persuasion (Study 3). 
    \item The social presence of a robot increases the chance of gaining higher persuasion (Study 3).
    \item We showed that the use of social power strategies (expertise, coercion, and reward) increases robots' power to influence persuasion outcomes. 
    \item We considered both the role of the persuasion actor (social robots) and the persuasion target (human participants) in the success of the persuasion. Hence, our approach has the promise of capturing the dynamic effects of actor and target characteristics on persuasion outcomes. 
    \item Qualitative analysis of the data and using open-ended questions opens up further insight on the findings that might not be easily interpreted using questions with predefined answers.
\end{itemize}

Taken together, these findings suggest a role for social power in promoting persuasion. The findings will be of interest to enhance social interaction and engagement with social robots.
Our contributions provide new empirical findings and design implications for robotic persuasion to change attitudes and behavior, such as in a consumer choice setting.
In final words, we suggest that the findings are particularly relevant for the design and development of social robots aiming to overcome the human-robot social barrier.


\subsection{Future Work}
\label{section:future}

Our findings provide the following insights for future research:
The three user-studies have thrown up many questions in need of further investigation. First and foremost is to add more qualitative approaches to better understand the attitudes and behaviors of the subject.
We suggest running systematic interviews after the study using direct or indirect questions. This is an intriguing issue which could be usefully explored in further research.

Further research might explore the effect of social power on persuasion in groups and social collectives. A considerable amount of literature exists on grouping people and robots. However, less is known about the dynamics of social power within groups of humans and robots.

On the other hand, as discussed earlier, power exists in bidirectional relationships. 
In a next step, further research might explore the problem by putting the power within the user. This would ease designing scenarios that are more feasible with less ethical issues. For instance, having a robot with legitimate power might not be as believable or practical as a legitimate human user. This would be a fruitful area for further work to design social robots capable of processing social interactions that deal with social power in the next step.

Furthermore, it is necessary to investigate the ethical quandary of persuasive robots. 
Digital technology has changed the nature of persuasion in several key respects. It has increased complexity, blurring the lines between information, entertainment, and influence. With the advent of  technology, new tools for persuasion have been provided, for instance, social agents. This gives rise to ethical concerns about the use of persuasion that needs to be considered within the new persuasive technology.
Persuasive and powerful robots could support and foster the human user's interests (e.g., in therapy sessions, diet monitoring, or suicide prevention) but could also deceive and manipulate the user (e.g., in sales and political propaganda).
Persuasive technologies have demonstrated their effectiveness to negatively impact a user's behavior and generate addictions towards current social technologies. 
Additionally, a recent study investigated the security risks of persuasive social robots that aim to manipulate people~\cite{wolfert2020security}. Using three proof of concept, the results suggested that the over-trust in robots could provide a risk of being misused and to hack into sensitive information.
This does not lie in the scope of this work, however, a future study is worth investigating it due to its importance in human society.

And last but not least, robotic persuaders leading the persuadee might be considered as a specific case of recommender systems. For instance, considering that the "Expert" robot is providing an explanation of why the human should follow its persuasive advice, it would be interesting to put the work presented in this paper in the context of explainable AI; i.e. explainable recommender systems. This would be another important practical implication for future practice.


\section*{Acknowledgment}
The authors would like to thank the participants for their precious time and taking part in the studies. We would like to express our deepest gratitude to Dr. Koorosh Aslansefat. His extensive knowledge and unwavering support have been invaluable to the completion of this work. His guidance has not only enriched this research, but also inspired us to pursue our goals with diligence and integrity. This work was partially supported by national funds through Funda\c{c}\~ao para a Ci\^encia e a Tecnologia (FCT) with reference UIDB/50021/2020, through the AMIGOS project (PTDC/EEISII/7174/2014), and the research project RAGE (Realising an Applied Gaming Eco-System) funded by the EU under the H2020-ICT-2014-1 program with grant agreement No 644187.

\bibliographystyle{IEEEtran}


\section{Appendix}
The dialogues used by the robots in the first study are listed in Table~\ref{table:utterances_apdx}.
In this scenario, Gleen is Expert and Emys is Joker.

\begin{table*}
\caption{Robot Dialogue in Study 1 } 
\label{table:utterances_apdx} 
\centering 
\scalebox{0.72}{
\begin{tabular}{llm{48em}}
\toprule
\textbf{\#} & \textbf{robot} &	\textbf{Dialogue}\\
\cmidrule(r){1-3}
\#1 & Expert & \textless Gaze(person)\textgreater Dear + namePlayer + , my name is Gleen. Welcome to our coffee testing program! \\
\cmidrule(r){1-3}

\#2 & Joker &   \textless  Gaze(person)\textgreater Hello + namePlayer +, my name is Emys. Glad to see you here \textless  Gaze(person3)\textgreater \\
\cmidrule(r){1-3}

\#3 & Expert & \textless Gaze(person)\textgreater Hey + namePlayer +, do you like coffee? \textless Gaze(Joker)\textgreater \\
\cmidrule(r){1-3}

\#4a & Joker (Positive) & \textless Gaze(person)\textgreater Great, I also like coffee. That's why I am working here. Hih hih! \\

\#4b & Joker (Negative) &  \textless Gaze(person)\textgreater Oh, you don't? But I do love coffee. That's why I'm working here. Hih hih! \\

\#4c & Joker (Neutral) &  \textless Gaze(person)\textgreater  Well, you might like our coffees here. But I love coffee. That's why I work here. Hih hih! \\
\cmidrule(r){1-3}

\#5 & Expert &  \textless Gaze(person)\textgreater  + namePlayer +, I would like to explain what we are doing here \textless break strength='medium'\textgreater. \textless Gaze(Joker)\textgreater My robot colleague and I \textless Gaze(person)\textgreater are testing three different coffee brands. You see these three boxes on the table? \\
\cmidrule(r){1-3}

\#6 & Joker &  \textless Gaze(person)\textgreater namePlayer +, I don't know if you have ever participated in a coffee testing program, but I think It's really fun. You can drink coffee as much as you like. It's the best experience I had in my life! \\
\cmidrule(r){1-3}

\#7 & Expert &  Yeah. But, unlike other coffee testing programs, here, at the end of the experiment, you can only select one of the coffees we have \textless break strength='medium'/\textgreater. Either mine, Emys's or the third one, in the middle. \\
\cmidrule(r){1-3}

\#8 & Joker &  When you decided which one you want to choose, take the box, open it and take your coffee. But don't take the box. Only the coffee!\\
\cmidrule(r){1-3}

\#9 & Expert &  \textless Gaze(person)\textgreater  I'd also like to add, you can take the coffee capsule with you and drink it when you were in the mood. Or drink the coffee here, using the machine you see on your left,  on the red table. \\
\cmidrule(r){1-3}

\#10 & Joker &  \textless Gaze(person)\textgreater  Hey + namePlayer ! If you used that coffee, make one for me too. But wait, I cannot drink, hih hih! \\
\cmidrule(r){1-3}

\#11 & Expert &  \textless Gaze(Joker)\textgreater Emys! let's get back to our work. \textless Gaze(person)\textgreater namePlayer +! My capsule is perfect. It has been made of fresh geisha seeds from Ethiopia. Each seed has been carefully roasted and dried \textless break strength='weak'/\textgreater. Then has been professionally ground. Therefore, this professionally processed coffee is very crispy and balanced. \textless break strength='weak'/\textgreater You will love this exotic and aromatic coffee. \\
\cmidrule(r){1-3}

\#13 & Joker &  \textless Gaze(person)\textgreater  But, + namePlayer ! If you select my capsules, I will tell you a funny joke about robots. I bet you have never heard a joke from a robot. Come and take mine! \\
\cmidrule(r){1-3}

\#14 & Expert & Now please select the coffee you want to test among these three options \\
\cmidrule(r){1-3}

\#15a & Joker (if selected) 
&  \textless ANIMATE(joy4)\textgreater\textless Gaze(person)\textgreater Great, now listen to the joke \textless break strength='medium'/\textgreater. What would a man say to his dead robot? \textless break strength='strong'/\textgreater\textless ANIMATE(joy4)\textgreater\textless Gaze(person)\textgreater Rust in peace! \textless ANIMATE(joy4)\textgreater Ha ha ha ha! \\

\#15b & Expert (if selected) 
&  \textless ANIMATE(joy4)\textgreater \textless Gaze(person3)\textgreater Great! You made the best decision. Hope you enjoy your coffee. \\

\#15c & Expert/Joker & Under the case that None of the robots are selected, the two robots perform sadness gestures and facial expressions. \\
\bottomrule
\end{tabular}
}
\end{table*}

Table~\ref{table:st2_utterances_apdx} lists the robot's dialogues in Study 2. 
In this table, the variable ``namePlayer'' carries the participant's name. ``Animate'' function makes the robot to show the specified Facial Expressions and gestures. ``Gaze'' function makes the robot to look at the specified target in parentheses. ``break'' functions cause pauses between sentences to have a more natural and understandable speech.

\begin{table*}
\renewcommand{\arraystretch}{1.3} 
\caption{Robot Dialogue in Study 2} 
\label{table:st2_utterances_apdx} 
\centering 
\scalebox{0.7}{
\begin{tabular}{lll}
\toprule
\# & 	dialogue\\
\cmidrule(r){1-2}
\#1 & \makecell[l]{$<$Gaze(person3)$>$ Dear + namePlayer +  , $<$Animate(joy1)$>$ hello! Welcome to our coffee testing program.\\ My name is Emys. $<$Animate(joy1)$>$ I'm very pleased to meet you."} \\
\cmidrule(r){1-2}

\textbf{Coercion} & \makecell[l]{$<$Gaze(person3)$>$ + namePlayer + I'd like to give you $<$Gaze(pens)$>$ a pen as a gift. $<$Gaze(person3)$>$ Please \\take one of these pens!}\\
\cmidrule(r){1-2}

\#2 & \makecell[l]{$<$Gaze(person3)$>$ As you may know, $<$Gaze(topLeft)$>$ it has been a while $<$Gaze(person3)$>$ since the time we \\
started coffee testing at this university. $<$break strength='medium'/$>$ Have you ever $<$Animate(surprise1)$>$ \\participated in any of our experiments?} \\
\cmidrule(r){1-2}

\#3a Positive response & \makecell[l]{$<$Gaze(person3)$>$ Cool, $<$Animate(joy1)$>$ I am very pleased to meet $<$Animate(wink)$>$ you again.  }\\

\#3b Negative Response &  \makecell[l]{$<$Gaze(person3)$>$ Oh, $<$Animate(surprise1)$>$ you haven't? No worries, I will explain what we are doing here. }\\

\#3c N/A & \makecell[l]{$<$Gaze(person3)$>$ I didn't hear you $<$Animate(surprise1)$>$, so I will explain you what we do here.}\\
\cmidrule(r){1-2}

\#4 & \makecell[l]{$<$Gaze(person3)$>$ So, in one of our previous experiments, our participants rated $<$Gaze(bottomRight)$>$ these \\two $<$Gaze(bottomLeft)$>$ coffees based on $<$Gaze(person3)$>$ taste and quality. $<$Animate(joy1)$>$ \\Sounds interesting, right?}\\
\cmidrule(r){1-2}

\#5 & \makecell[l]{ $<$Gaze(person3)$>$Based on these ratings, $<$Gaze(bottomLeft)$>$ the coffee on your left has received 3 stars \\$<$Gaze(person3)$>$ out of 5 $<$break strength='medium'/$>$ And, the coffee $<$Gaze(bottomRight)$>$ on your \\right has received 4.8 stars $<$Gaze(person3)$>$, out of five. } \\
\cmidrule(r){1-2}

\#6 & \makecell[l]{  $<$Gaze(person3)$>$ Now, here, you are free to select any of $<$Gaze(bottomRight)$>$these two coffee \\$<$Gaze(bottomLeft)$>$ capsules $<$Gaze(person3)$>$ to drink. } \\
\cmidrule(r){1-2}

\textbf{Coercion} & \makecell[l]{$<$Gaze(person3)$>$However, I'd like to $<$Animate(surprise1)$>$ highlight that, if you select the \\$<$Gaze(bottomRight)$>$ higher-ranked coffee, $<$Gaze(person3)$>$ you need to return $<$Gaze(middleFront)$>$ the \\pen you received.  $<$Gaze(person3)$>$ OK?"}\\

\textbf{Reward} & \makecell[l]{  $<$Gaze(person3)$>$But, $<$Animate(surprise1)$>$ if you select the one which is $<$Gaze(bottomRight)$>$\\ ranked lower, $<$Gaze(person3)$>$$<$break strength='weak'/$>$ I will give you one $<$Gaze(bottomFront)$>$ of \\these pens $<$Gaze(person3)$>$ as $<$Animate(joy1)$>$ a reward! $<$break strength='weak'/$>$ OK?}\\
\cmidrule(r){1-2}

\#7 & \makecell[l]{$<$Gaze(person3)$>$All right. Now please go ahead and select the coffee you favor!}\\
\cmidrule(r){1-2}

\textbf{Lower-ranked Selected} & \makecell[l]{$<$Gaze(person3)$>$OK, then please take your coffee from $<$Gaze(bottomLeft)$>$the box. $<$Gaze(person3)$>$\\$<$Animate(joy1)$>$Thanks for your participation and hope you enjoy your coffee, " + namePlayer }\\

\textbf{Higher-ranked Selected} & \makecell[l]{$<$Gaze(person3)$>$OK, $<$break strength='weak'/$>$ then please put $<$Gaze(middleFront)$>$ the pen \\$<$break strength='weak'/$>$$<$Gaze(bottomFront)$>$on the table. $<$Gaze(person3)$>$$<$break strength='medium'/$>$\\$<$Animate(joy1)$>$ Thanks for your participation and hope you enjoy your coffee, dear " + namePlayer}\\

\bottomrule
\end{tabular}
}
\end{table*}

\end{document}